\documentclass[useAMS,usenatbib]{mn2e}
\usepackage{graphicx}
\usepackage{natbib}
\usepackage{longtable}

\title[The GRB 060505 host in high resolution]{The host of the SN-less GRB 060505 in high resolution\thanks{Based on observations under ESO 081.D-0569(A)}}
\author[C. C. Th\"one et al.]{C. C. Th\"one$^{1}$\thanks{cthoene@iaa.es}, L. Christensen$^{2}$, J. X. Prochaska$^{3}$, J. S. Bloom$^{4}$, J. Gorosabel$^{1,5,6}$ \and J. P. U. Fynbo$^{2}$, P.  Jakobsson$^{7}$, A. S. Fruchter$^{8}$\\
$^{1}$Instituto de Astrof\'isica de Andaluc\'ia, Glorieta de la Astronom\'ia s/n, 18008 Granada, Spain\\
$^{2}$DARK Cosmology Centre, Niels-Bohr-Institute, University of Copenhagen, Juliane Maries Vej 30, 2100 K\o benhavn \O, Denmark\\
$^{3}$Dep. of Astronomy and Astrophysics \& UCO/Lick Observatory, Univ. of California, 1156 High Street, Santa Cruz, CA 95064, USA\\
$^{4}$Dep. of Astronomy, University of California, Berkeley CA, 94720-3411, USA\\
$^{5}$Unidad Asociada Grupo Ciencia Planetarias UPV/EHU-IAA/CSIC, Dep. de F\'{\i}sica Aplicada I, E.T.S. Ingenier\'{\i}a, Universidad del\\
~\,Pa\'{\i}s Vasco UPV/EHU, Alameda de Urquijo s/n, E-48013 Bilbao, Spain.\\
$^{6}$Ikerbasque, Basque Foundation for Science, Alameda de Urquijo 36-5, E-48008 Bilbao, Spain.\\
$^{7}$Centre for Astrophysics and Cosmology, Science Institute, University of Iceland, Dunhagi 5, 107 Reykjav\'ik, Iceland\\
$^{8}$Space Telescope Science Institute, 3700 San Martin Drive, Baltimore, Maryland 21218, USA
}

\begin{document}

\date{Accepted. Received; in original form }

\pagerange{\pageref{firstpage}--\pageref{lastpage}} \pubyear{2014}

\maketitle

\label{firstpage}

\begin{abstract}
{The spiral host galaxy of GRB 060505 at $z=0.089$ was the site of a puzzling long duration burst without an accompanying supernova. Studies of the burst environment by Th\"one et al. (2008) suggested that this GRB came from the collapse of a massive star and that the GRB site was a region with properties different from the rest of the galaxy. We reobserved the galaxy in high spatial resolution using the VIMOS integral-field unit (IFU) at the VLT with a spaxel size of 0.67 arcsec. Furthermore, we use long slit high resolution data from HIRES/Keck at two different slit positions covering the GRB site, the center of the galaxy and an HII region next to the GRB region. We compare the properties of different HII regions in the galaxy with the GRB site and study the global and local kinematic properties of this galaxy. The resolved data show that the GRB site has the lowest metallicity in the galaxy with $\sim$1/3 Z$_\odot$, but its specific SFR (SSFR) of 7.4 M$_\odot$/yr/L/L* and age (determined by the H$\alpha$ EW) are similar to other HII regions in the host. The galaxy shows a gradient in metallicity and SSFR from the bulge to the outskirts as it is common for spiral galaxies. This gives further support to the theory that GRBs prefer regions of higher star-formation and lower metallicity, which, in S-type galaxies, are more easily found in the spiral arms than in the centre. Kinematic measurements of the galaxy do not show evidence for large perturbations but a minor merger in the past cannot be excluded. This study confirms the collapsar origin of GRB060505 but reveals that the properties of the HII region surrounding the GRB were not unique to that galaxy. Spatially resolved observations are key to know the implications and interpretations of unresolved GRB hosts observations at higher redshifts.
}

\end{abstract}


\section{Introduction}

GRB 060505 and its host galaxy have drawn particular attention in the GRB community for several reasons. GRB 060505 was one of two nearby long-duration GRBs that had no accompanying supernova \citep{Fynbo06, Gehrels06, DellaValle06, GalYam06}. Due to its duration of only $\sim$ 4\,s it was suggested that it could have intact been a short GRB that naturally has no SN \citep{Ofek06}. However, the presence of a significant spectral lag \citep{McBreen06} and its afterglow properties \citep{Xu09} favour a long GRB origin. The host galaxy was somewhat unusual for a long-duration GRB, a late-type, moderately star-forming, solar metallicity spiral galaxy. \cite{Thoene08}, however, found that the properties of the stellar population at the GRB site very much resembled the global properties of dwarf GRB hosts \citep[e.g.][]{Christensen04, Savaglio09,Hjorth13} concerning metallicity and star formation rate. Furthermore, it had a very young age, making the connection to a massive star even more likely.

GRB progenitor models suggest that long GRBs can only be produced by metal poor stars, otherwise, large mass losses of the massive progenitor slows down the star such that it cannot form an accretion disk at collapse \citep[e.g.][]{Woosley06}. Indeed, most long-duration GRBs have been found in metal poor environments or galaxies though a few exceptions for high metallicity environments have been found, e.g. GRB 020819 \citep{Levesque10a}, GRB 050826 \citep{Levesque10b}, GRB 080605 \citep{Kruehler12} and GRB 090323 \citep{Savaglio12}. Long duration GRB hosts furthermore show a moderate to high star formation rate. In the low redshift universe those conditions are mainly met in dwarf irregular and blue compact galaxies, but probably also in the outskirts of late-type galaxies. Short GRB hosts seem to show different properties \citep[e.g.][]{Berger13} in line with the suggested origin in the coalescence of two compact objects.

In recent years, spatially resolved studies of GRB and SN hosts have become increasingly important. High spatial resolution imaging with the {\it HST} correlated the brightness in the blue at different parts in the host to the one at the explosion site. While SN Type II are distributed evenly, GRBs and Type Ic SNe are associated to the blues regions of their hosts \citep{Fruchter06, Kelly08}. Likewise, WC (cabon-oxygen Wolf-Rayet) stars show a spatial distribution similar to SNe Type Ic and GRBs concerning the brightness of their surrounding region while WN (nitrogen Wolf-Rayet) stars correlate with the distribution of SNe Ib \citep{Leloudas10}. Short GRBs, in contrast, are found predominantly in dimmer regions and with larger offsets from their host galaxy, consistent with an older population and/or kicks of the progenitor system from the birth site by the preceding SN explosion \citep{Fong10}.
 
Following the mass-metallicity relation, some more massive GRB hosts show a higher global metallicity of solar or supersolar values. However, most of them have actually a lower metallicity at the location of the GRB \citep{Levesque10a, Graham13}. As they are usually located in the outskirts of their hosts this is expected due to the usual metallicity gradient in star-forming galaxies. However, some GRB sites have still higher metallicities than theoretical models would require. GRB 0208019 \citep{Levesque10a} f.ex. showed a super-solar metallicity at the GRB site, however, it was a ``dark GRB'' (no optical counterpart was detected \citet{Jakobsson020819}). It is still debated weather the high metallicity might be the reason that those dark GRBs are intrinsically faint \citep[see e.g.][]{Graham13} and whether we have to revise our theoretical models. The site of the short GRB 080905A \citep{Rowlinson} in contrast showed an old population with little star formation, as expected.

Comparing GRBs and SNe, \cite{Modjaz08} found a lower metallicity for GRB sites compared to broad-line Type Ic SNe without GRB. The situation is less clear for comparing different types of SNe, while \cite{Modjaz11} finds a lower metallicities for Type Ib SNe compared to Type Ic, \cite{Anderson10, Leloudas12} do not see significant differences. Recent IFU studies of SN sites indeed show differences in metallicity, SFR and age between different types of SN \citep{Ku13a, Kun13, Galbany13}. 

Detailed resolved studies using integral-field-spectroscopy has to date only been performed for the host of GRB 980425, the first GRB observed to be connected to a broad-line Type Ic SN \citep{Christensen08}. The GRB site turned out to be not the youngest, most metal poor and star-forming region in the galaxy and did not show any WR features, in contrast to another star-forming region next to the one of the GRB site. This might imply that GRBs do not need the extreme properties as suggested by stellar population models and/or other evolutionary channels such as binaries might play an additional role. However, the sample of well studied GRB environment is still somewhat small to get some clear indications on the progenitors of these interesting events.

In this paper, we present integral field spectroscopic data of the host galaxy of GRB 060505 together with high resolution spectra in  two slitpositions placed along the galaxy (Sec. 2). This allows us to study the properties of the different star-forming regions in the galaxy in terms of metallicity, star-formation rate, stellar population age and extinction and to compare the GRB region with other HII regions in the galaxy (Sec. 3). Furthermore, we investigate the kinematics of the galaxy to search for possible perturbations and other reasons for an increase in star-formation at the GRB site (Sec. 4).

\section{Observations}\label{observations}
\begin{figure*}
\includegraphics[width=13.5cm]{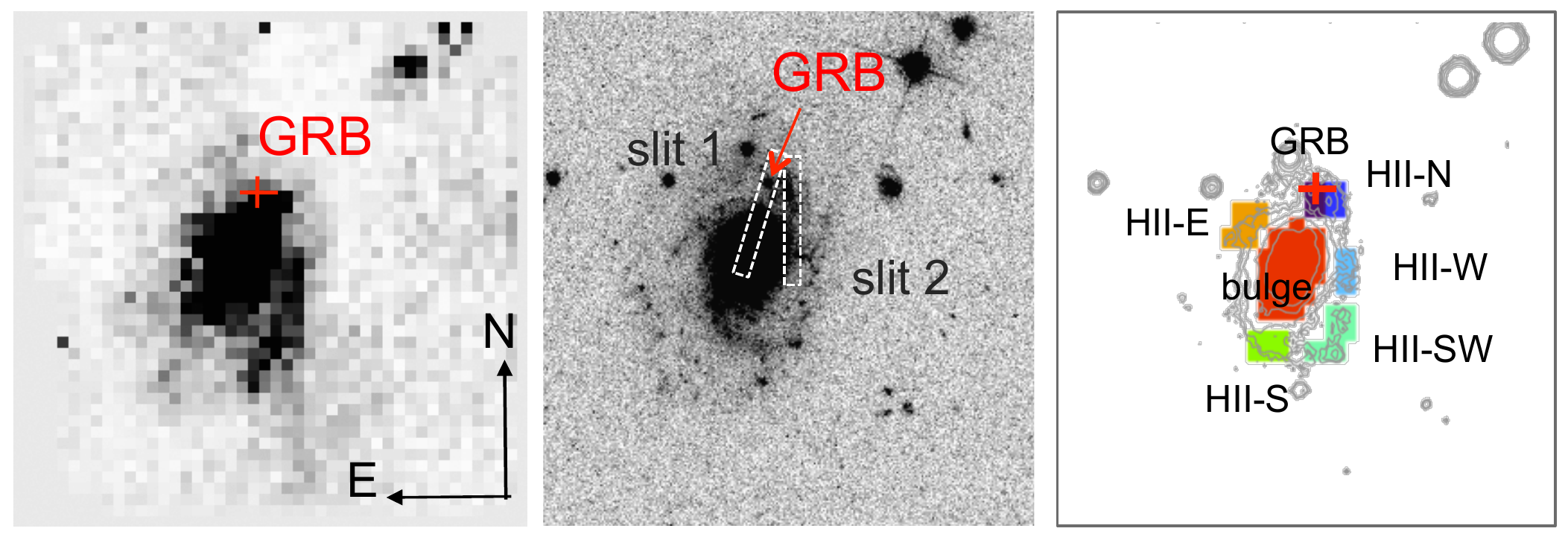}
\caption{Left panel: Image of the host galaxy of GRB 060505 in H$\alpha$. Middle panel: Positioning of the two HIRES slits overlaid on an image from HST/ACS in the F814W filter. Right panel: Contours of a FORS/VLT V-band image (same as in the following figures) and the different regions extracted as integrated spectra from the VIMOS datacube. The naming of the regions is used througout the paper. The site of the GRB is indicated with a cross or an arrow in the three panels, FOV in all three panels is $\sim$ 30''$\times$30''.
\label{VIMOSslits}}
\end{figure*}

We obtained
9.85\,h of observations with the integral field unit (IFU) of the VIMOS instrument at the VLT between April and August 2008. The setting with 0.67'' lenslets and a 27'' $\times$ 27'' field-of-view (FOV) provides a sampling of $\sim$~1\,kpc and easily covers the entire galaxy with one pointing. The seeing conditions during most of the observations, however, were larger than 0\farcs7, reducing the resolution to 1\farcs34 or 2.2\,kpc. Three different gratings were used, HR orange, HR blue and LR blue, to cover the wavelength range from 4000 to 7400 \AA{} which includes all important diagnostic emission lines from [OII] to [NII]. The LR blue grating provides a resolution of $\lambda/\Delta\lambda$=250 whereas the HR orange and blue grism have resolutions of 2700 and 2600 respectively which barely resolves the emission lines.

The data were reduced and fluxcalibrated using the ESO VIMOS pipeline. The background was subtracted taking blank sections outside the galaxy in each of the four quadrants of the chip since they have slightly different readout noise and background levels. For each grism, all exposures were stacked to one single datacube and bad fibres masked out before the combination. The absolute flux calibration was scaled to match the broad-band magnitudes as listed in \cite{Thoene08} using the integrated spectrum of the galaxy which also accounts for the large relative offset between the LRblue and the HR grisms. Fig. \ref{VIMOSslits} shows an ``image'' of the galaxy summed in wavelength over the width of the H$\alpha$ line.

Furthermore, on July 16, 2007, we obtained two spectra with exposure times of 1800 and 1200s with HIRES and Keck I at the Keck Observatory on Mauna Kea, Hawaii. Using a 0\farcs8 slit, HIRES provides a spectral resolution of 45,000 or $\sim$ 7 km/s. With the red setting used, the orders almost overlap in the blue between 3990 and 4100 \AA{} but this does not affect the [OII] emission lines present in this part of the spectrum. The HIRES slit has a length of 7'' and was positioned at two different angles on the galaxy to cover the HII region at the GRB site, the center of the galaxy and another nearby HII region. The first position was chosen at the same angle (17 deg) as for the low-resolution spectra presented in \cite{Thoene08}, the second slit was positioned at zero degrees or North-South to cover the other large HII region west of the GRB position (see Fig.\ref{VIMOSslits}). 

\begin{figure}
\centering
\includegraphics[height=6.6cm]{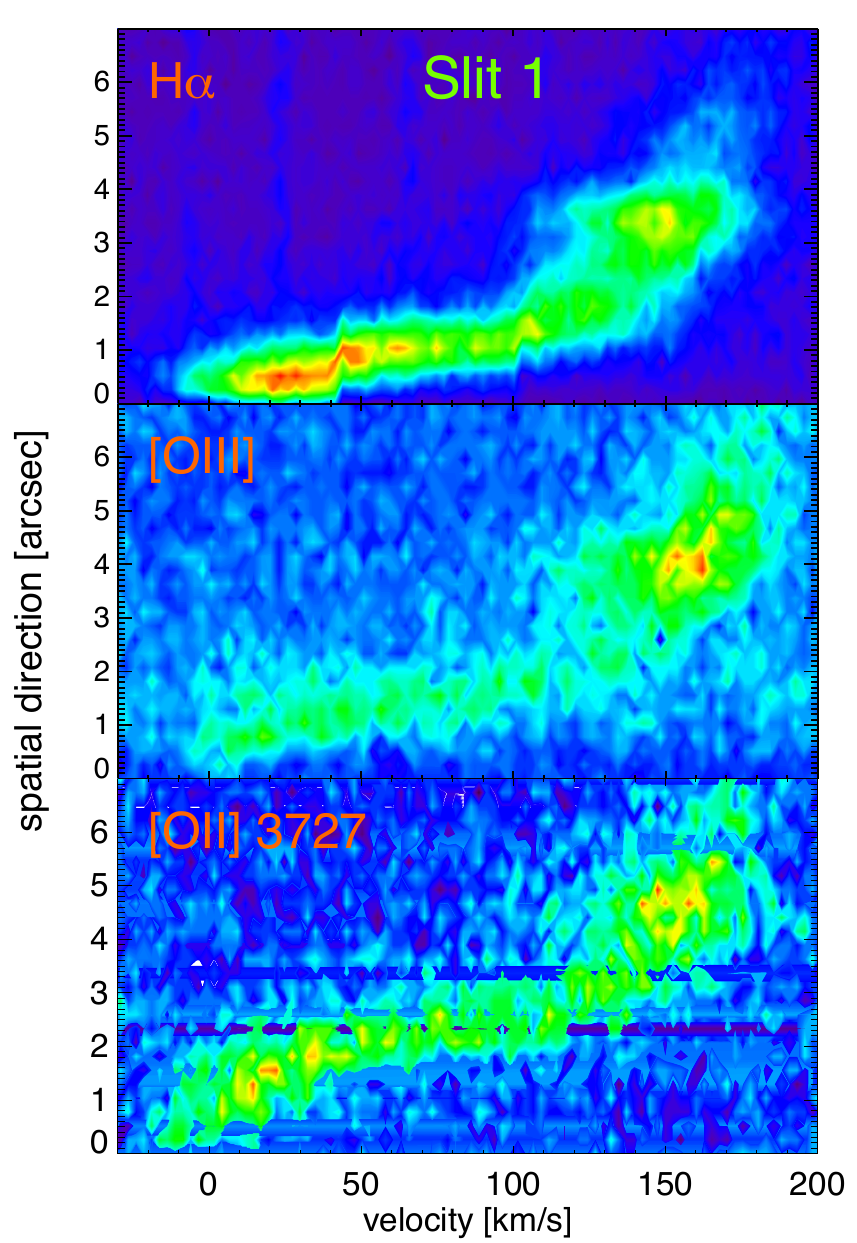}\hspace{-4mm}
\includegraphics[height=6.535cm]{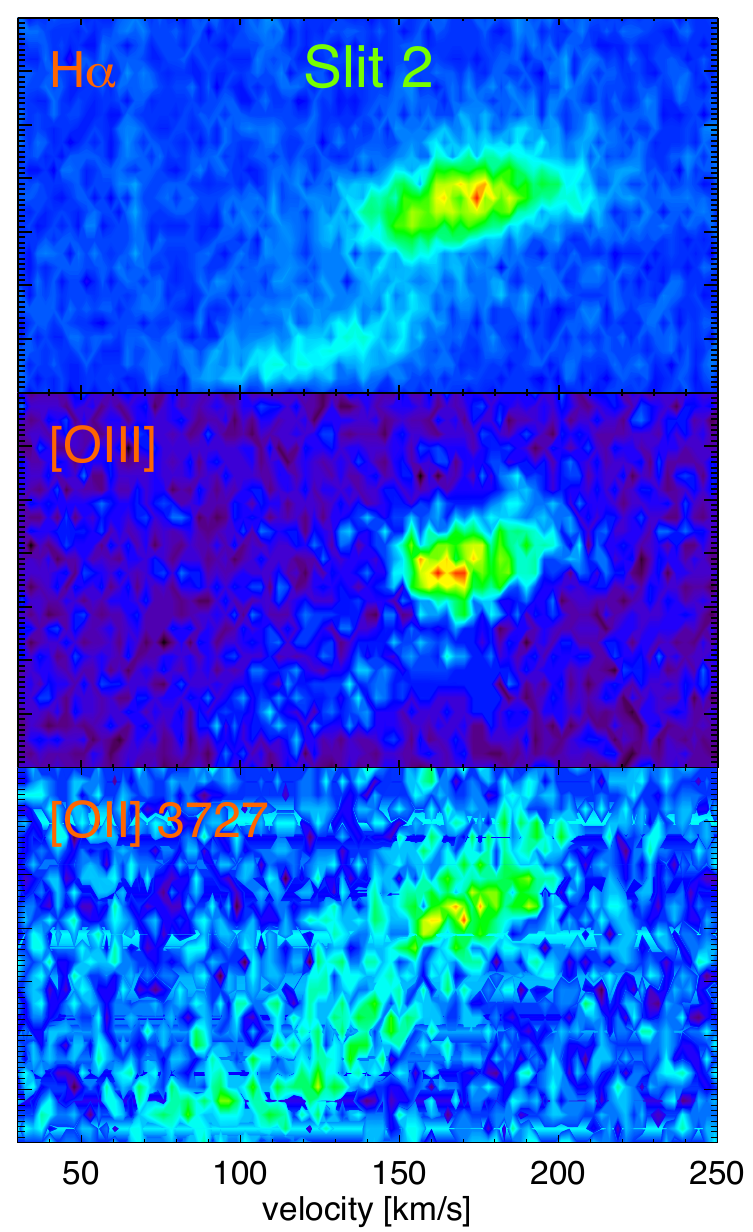}
\caption{Contour plots for H$\alpha$, [O{\sc iii}] and [O{\sc ii}] along the two slit positions, left: along the major axis (slit 1), right: through the HII region next to the GRB site (slit 2). v=0 km/s corresponds roughly to the center of the galaxy. The color scale is a linear distribution of the pixel values for each individual line (not calibrated in absolute fluxes). \label{060505:hiresplots}}
\end{figure}

The Keck/HIRES data were reduced using the HIRedux pipeline\footnote{http://www.ucolick.org/$\sim$xavier/HIRedux/index.html}
bundled within the XIDL\footnote{http://www.ucolick.org/$\sim$xavier/IDL/index.html} software package.  We used the pipeline to produce two-dimensional
spectral images, calibrated in wavelength by tracing the ThAr lamp exposures.  A rough estimate of the flux calibration was performed using the 1D spectra of spectrophotometric standard stars obtained on the same night.  These do not account for slit losses but should provide good estimates ($\approx 10\%$) to the relative flux. Fig. \ref{060505:hiresplots} shows cut-outs of the 2D spectral images where $v=0$\,km/s
corresponds to the velocity recorded at the approximate center of the
host galaxy.  A significant shear in the velocity field is evident
with increasing offset, indicating rotation of the galaxy. 

\section{Resolved properties}
The host of GRB 060505 is a low-mass (7.9$\times$10$^{9}$ M$_\odot$), Sbc spiral galaxy seen almost face on with an inclination of 49 degrees \citep{Thoene08}. The GRB happened in a star-forming (SF) knot in the northern spiral arm of the galaxy. As shown in Fig. \ref{VIMOSslits}, just west of the GRB region is another SF region (named ``HII-N''). The host contains a number of other SF regions in the spiral arm to the south and west (HII-W) and south-west (HII-SW), south of the bulge (HII-S) and SF regions in the spiral arm to the east (HII-E). In the following, we apply the naming convention as marked in Fig. \ref{VIMOSslits}. 

\begin{figure*}
\centering
\includegraphics[width=15cm]{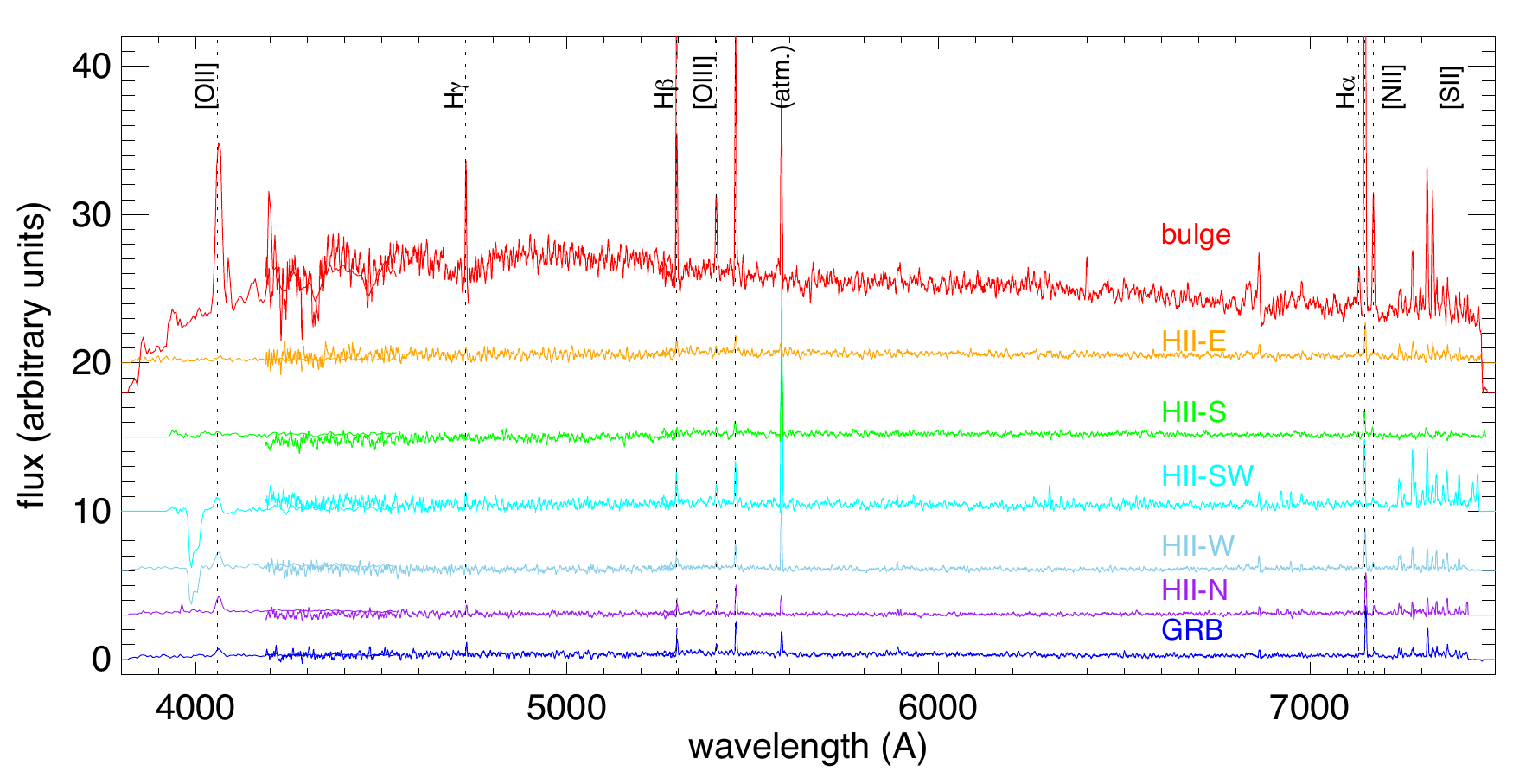}
\caption{Integrated spectra of the different regions in the VIMOS dataset indicated in Fig. \ref{VIMOSslits}, offset by an arbitrary value for clarity. The spectra show all the three grisms used, the wavelength range was cutted in the overlapping wavelength regions and the fluxes scaled such as to exactly match the different spectra. \label{060505:specsregions}}
\end{figure*}

From the VIMOS/IFU data we create maps of emission lines and different properties across the host galaxy. To produce the emission lines maps we sum the flux of 15 pixels around the center of the corresponding emission line in each spaxel. The line continuum was subtracted taking the average of $\sim$10 pixels in the vicinity of the emission line free of other features. In the individual spaxels, only the LR blue grism shows some continuum emission due to its higher S/N per pixel. We furthermore extract integrated spectra for the different regions denoted in Fig. \ref{VIMOSslits} and measure line-fluxes with standard tasks in IRAF, the integrated spectra are plotted in Fig. \ref{060505:specsregions}.

In the spectra of the individual spaxels we detect all of the strong emission lines of [OII], [OIII], H$\beta$, H$\alpha$ and [NII]. [SII] and H$\gamma$ are only detected in the integrated spectrum of the bulge, the [SII] doublet furthermore lies in a region of strong atmospheric emission lines and could therefore not be used. Line fluxes and properties of the integrated spectra are given in Tab. \ref{fluxtable} and \ref{comptable}. A table with emission line fluxes and properties of all spaxels with fluxes of H$\alpha >$ 1.5$\times$10$^{-17}$ erg\,cm$^{-2}$\,s$^{-1}$ can be found in Tab. \ref{table:VIMOSfluxtable} in the appendix.

The same lines were found in slit 1 of the HIRES data. Due to higher sensitivity in the red and high resolution the [SII] doublet is clearly detected here, albeit affected by atmospheric emission lines, and we also detect [NII] $\lambda$ 6548. For slit 2 we only see the [OII] doublet, [OIII] $\lambda$ 5007, H$\beta$ and H$\alpha$, [NII] $\lambda$ 6583 and [SII] $\lambda$ 6717 are very faint [NII] $\lambda$ 6548 and [SII] $\lambda$ 6731 are completely absent.

The longslit HIRES data are divided into spatial bins at a resolution similar to the seeing of around 0\farcs5, corresponding to a physical dimension of $\sim$~0.8 kpc, from which we extract 1D spectra. We then fit Gaussian profiles to the emission lines in the extracted spectra, which represent a good fit to the emission lines in the individual spectra and determine their fluxes, center position and FWHM. The properties derived from this are found in Tab. \ref{table:HIRESval} in the appendix, due to the uncertainty in the absolute flux calibration of the HIRES data, we do not list the fluxes.

\begin{table*}
\begin{tabular}{l|llllllll}\hline
Region & [O\sc{ii}] & H$\beta$ & [O\sc{iii}] & [O\sc{iii}] & H$\alpha$ & [N\sc{ii}] & [S\sc{ii}] & [S\sc{ii}] \\
 & $\lambda$3727/29 & $\lambda$ 4862 & $\lambda$ 4960 & $\lambda$ 5008 & $\lambda$ 6564 & $\lambda$ 6585 & $\lambda$ 6718 & $\lambda$ 6732 \\ \hline\hline
GRB  &   2.39 $\pm$   0.13 &   1.15 $\pm$   0.07  &   0.76 $\pm$   0.08  &   1.98 $\pm$   0.06 &   3.11 $\pm$   0.07 &   0.24 $\pm$   0.03 &   1.75 $\pm$   0.07  &   1.36 $\pm$   0.47\\
HII-N  &   3.49 $\pm$   0.25 &   0.78 $\pm$   0.08  &   0.66 $\pm$   0.07  &   1.93 $\pm$   0.07 &   2.49 $\pm$   0.05 &   0.70 $\pm$   0.05 &  -0.20 $\pm$  -0.20  &   0.79 $\pm$   0.22\\
HII-W  &   1.27 $\pm$   0.42 &   1.01 $\pm$   0.22  &   0.44 $\pm$   0.04  &   1.34 $\pm$   0.04 &   2.70 $\pm$   0.06 &   0.46 $\pm$   0.08 &  -0.20 $\pm$  -0.20  &  -0.20 $\pm$  -0.20\\
HII-SW  &   3.08 $\pm$   0.50 &   1.56 $\pm$   0.23  &   1.45 $\pm$   0.13  &   2.83 $\pm$   0.09 &   4.16 $\pm$   0.10 &   0.35 $\pm$   0.04 &  -0.20 $\pm$  -0.20  &  -0.20 $\pm$  -0.20\\
HII-S  &   1.14 $\pm$   0.48 &   1.11 $\pm$   0.48  &   0.39 $\pm$   0.08  &   1.06 $\pm$   0.06 &   1.57 $\pm$   0.08 &   0.43 $\pm$   0.04 &  -0.20 $\pm$  -0.20  &  -0.20 $\pm$  -0.20\\
HII-E  &   1.21 $\pm$   0.31 &   1.16 $\pm$   0.15  &   0.49 $\pm$   0.20  &   0.88 $\pm$   0.07 &   2.19 $\pm$   0.09 &   0.19 $\pm$   0.03 &  -0.20 $\pm$  -0.20  &  -0.20 $\pm$  -0.20\\
bulge  &  36.82 $\pm$   4.06 &   8.86 $\pm$   0.26  &   9.49 $\pm$   0.35  &  16.30 $\pm$   0.35 &  37.20 $\pm$   0.91 &   9.18 $\pm$   0.17 &  10.13 $\pm$   0.36  &   9.78 $\pm$   0.66\\ \hline
galaxy  &  61.92 $\pm$   3.12 &  18.44 $\pm$   0.73  &  12.98 $\pm$   1.12  &  35.88 $\pm$   1.44 &  77.10 $\pm$   8.08 &  15.54 $\pm$   1.63 &  20.28 $\pm$   2.02  &  13.96 $\pm$   3.78\\
\hline\hline
\end{tabular}
\caption{Fluxes measured in the integrated spectra of different regions in the VIMOS dataset in units of 10$^{-16}$\,erg\,cm$^{-2}$\,s$^{-1}$ (see Fig.\ref{VIMOSslits}).  
} 
\label{fluxtable}
\end{table*}

\begin{table*}
\begin{tabular}{llllllll}\hline
 Region & 12+log(O/H) & 12+log(O/H) & log U &  E(B--V) & SFR & SSFR & EW H$\alpha$ \\
 & (N2) & (O3N2) &  & [mag] & M$_\odot$/yr & M$_\odot$/yr/L/L*/spaxel & \AA{}\\ \hline\hline 
GRB &   8.23 $\pm$   0.03 &    8.24 $\pm$   0.00 &  -3.08 $\pm$   0.04 &     0.00 &  0.047$\pm$  0.001 &   7.43 $\pm$   0.18 &     -57 $\pm$    5.9\\
HII-N &   8.49 $\pm$   0.02 &    8.33 $\pm$   0.01 &  -3.22 $\pm$   0.05 &     0.14 &  0.038$\pm$  0.001 &   7.37 $\pm$   0.16 &    -130 $\pm$   13.8\\
HII-W &   8.39 $\pm$   0.04 &    8.34 $\pm$   0.01 &  -3.00 $\pm$   0.16 &     0.00 &  0.041$\pm$  0.001 &   6.88 $\pm$   0.16 &     -83 $\pm$    9.2\\
HII-SW &   8.24 $\pm$   0.03 &    8.25 $\pm$   0.01 &  -3.05 $\pm$   0.09 &     0.00 &  0.063$\pm$  0.002 &  21.22 $\pm$   0.51 &     -47 $\pm$    5.4\\
HII-S &   8.48 $\pm$   0.03 &    8.42 $\pm$   0.04 &  -3.04 $\pm$   0.21 &     0.00 &  0.024$\pm$  0.001 &   6.07 $\pm$   0.32 &     -37 $\pm$    5.0\\
HII-E &   8.25 $\pm$   0.04 &    8.33 $\pm$   0.01 &  -3.13 $\pm$   0.15 &     0.00 &  0.033$\pm$  0.001 &   8.97 $\pm$   0.38 &     -21 $\pm$    3.9\\
bulge &   8.46 $\pm$   0.01 &    8.35 $\pm$   0.01 &  -3.30 $\pm$   0.06 &     0.40 &  0.566$\pm$  0.014 &   4.36 $\pm$   0.11 &     -38 $\pm$    4.0\\ \hline
galaxy &   8.42 $\pm$   0.04 &    8.32 $\pm$   0.01 &  -3.21 $\pm$   0.04 &     0.40 &  1.173$\pm$  0.123 &   6.49 $\pm$   0.68 &     -62 $\pm$    1.9\\
\hline\hline
\end{tabular}
\caption{Properties derived from integrated spectra in different regions  in the VIMOS dataset (see Fig.\ref{VIMOSslits}). Errors for the metallicity only includes the errors from the emission line fitting which are always smaller than the errors of the different calibrator (0.16\,dex for the N2 and 0.18\,dex for O3N2). log U is the ionization parameter obtained from [OII] to [OIII] as defined in \citealt{Diaz00}.} 
 \label{comptable}
\end{table*}

\subsection{Metallicity}\label{sect:metal}
The earlier longslit study on the host of GRB 060505 \citep{Thoene08} found a considerably lower metallicity (1/5 Z$_\odot$) at the site of the GRB compared to the other parts of the galaxy that have nearly solar values. With the VIMOS/IFU data we are now able to map the metallicity of the entire galaxy including the GRB site and determine whether the burst site is particularly metal poor or follows the general metallicity gradient in spiral galaxies \citep[e.g.][]{HW99}. 

The metallicity is determined by the N2 and O3N2 indices using a recent recalibration by \cite{Marino13} with systematic errors of 0.16 and 0.18\,dex respectively: 12+log(O/H)$=$8.743\,+\,0.462\,\,log$_{10}$(${\mathrm{[NII] 6583}}/\mathrm{H\alpha}$) for the N2-index and 12+log(O/H)$=$8.533\,--\,0.214 log$_{10}$~($\left( \mathrm{[OIII] 5007}/\mathrm{H\beta})*(\mathrm{H\alpha}/{\mathrm{[NII] 6583}})\right)$ for the O3N2 index. These indices are calibrated against metallicities derived from electron temperature measurements in nearby galaxies. For metallicities up to around solar values, those methods show a linear behaviour and are therefore a useful tool if direct measurements by the electron temperature are not available. Neither in the VIMOS nor in the HIRES data the temperature sensitive [OIII] $\lambda$ 4363 line is detected and we therefore have to rely on the indirect metallicity measurements. As solar metallicity we take 12+log(O/H)=8.69 \citep{Asplund09}. 

\begin{figure}
\includegraphics[width=\columnwidth, angle=0]{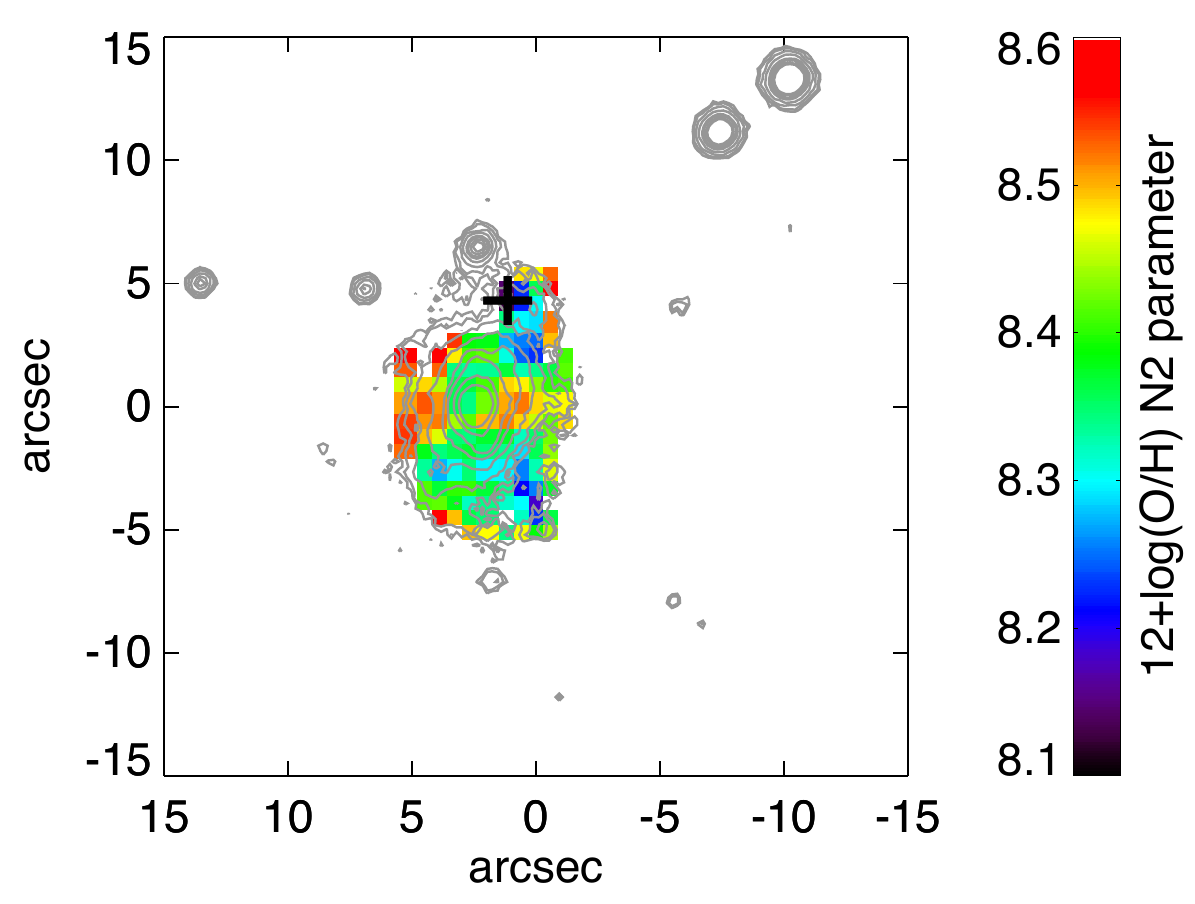}
\caption{Metallicity map from the VIMOS data using the N2 index, as solar value we use 12+log(O/H)=8.69 \citep{Asplund09}.
\label{N2}}
\end{figure}

\begin{figure}
\includegraphics[width=\columnwidth, angle=0]{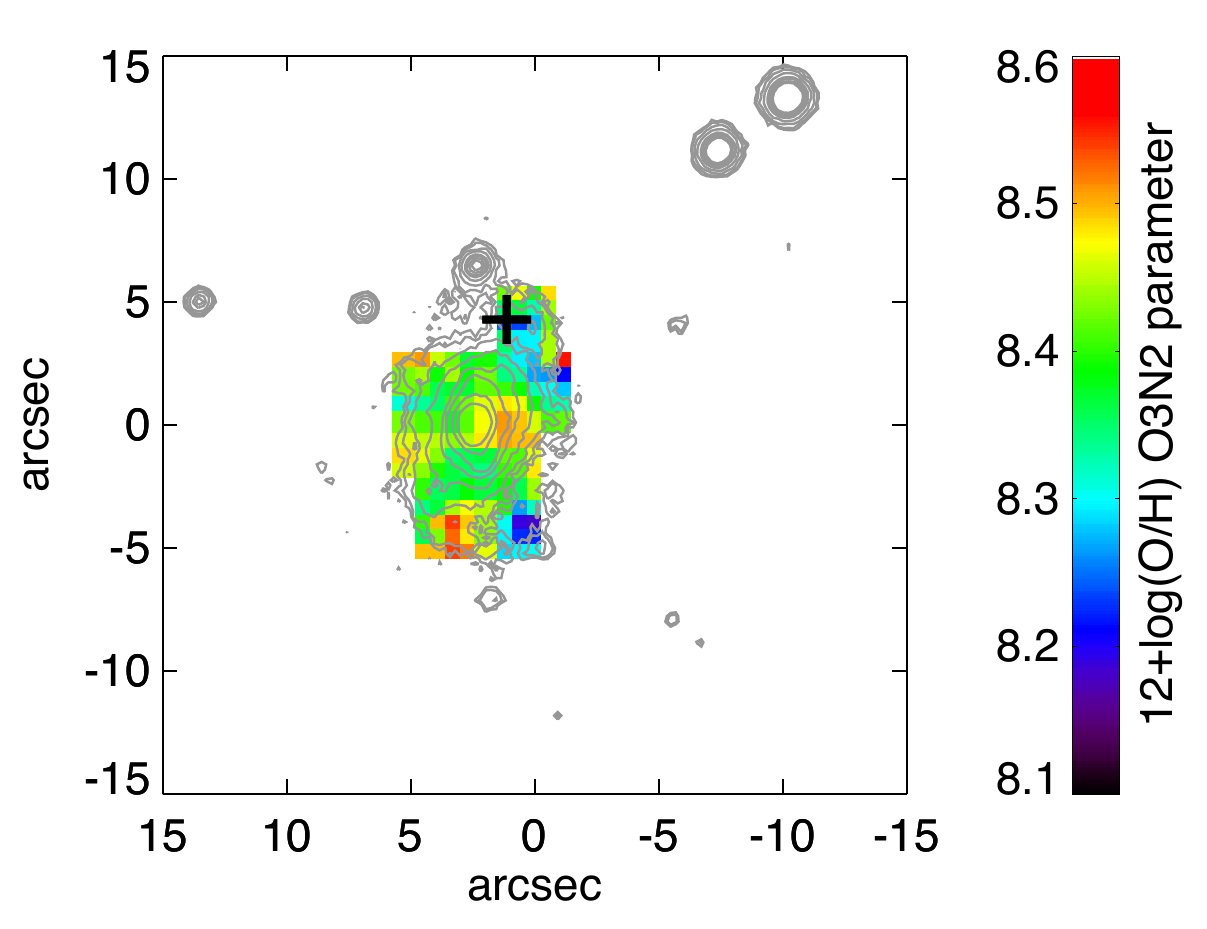}
\caption{Metallicity map using the O3N2 index, as solar value we use 12+log(O/H)=8.69 \citep{Asplund09}.
\label{O3N2}}
\end{figure}

The maps for metallicities determined via the N2 and O3N2 indices are shown in Fig. \ref{N2} and \ref{O3N2}. The VIMOS data give similar values for the metallicity determined from both methods. As expected in a star-forming spiral galaxy, there is a negative metallicity gradient from the centre to the spiral arms. Although the GRB region is the one with the lowest metallicity, similar values are reached in the SF regions in the SW spiral arm. The highest metallicities are actually not in the bulge but in a region just west of it. In the integrated spectra we get values of 12+log(O/H)$=$8.23 and 8.24 ($\sim$ 1/3 Z$_\odot$) from the N2 and O3N2 indices, somewhat higher than what has been previously determined from longslit data \citep{Thoene08}. Again, the SW spiral arm has similar metallicities whereas the SF region (HII-N) next to the GRB region has somewhat higher metallicities. The bulge has higher metallicities but is still subsolar, somewhat lower than what had been found in the longslit data.

\subsection{Ionization parameter}
\begin{figure}
\includegraphics[width=\columnwidth, angle=0]{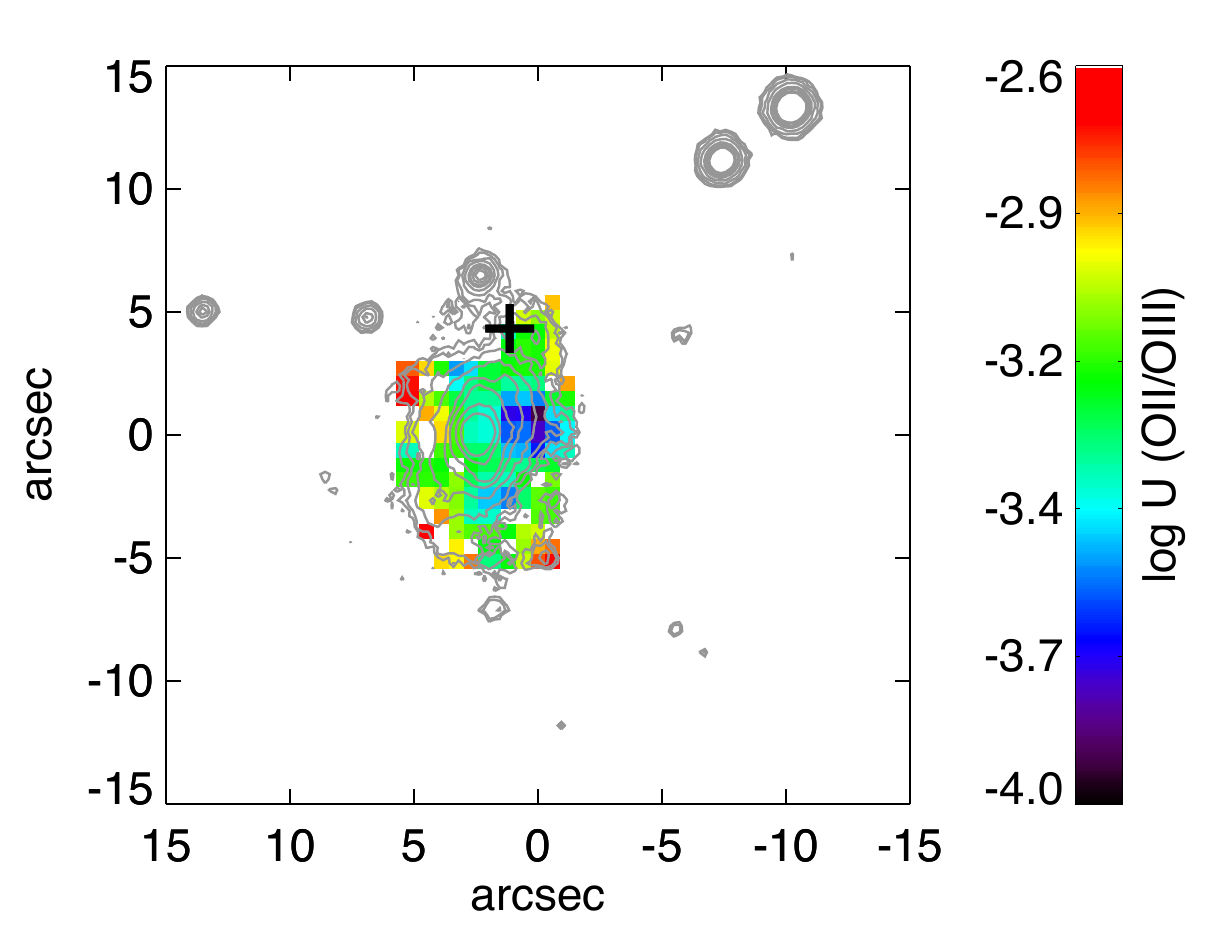}
\caption{Ionization measured via the ratio of [OII] to [OIII]. We use the ionization parameter U defined in \protect\cite{Diaz00}. 
\label{ionization} }
\end{figure}

Regions with ongoing star-formation normally also have a higher ionization. The ionization of the medium is important for the determination of metallicity since the ionization influences the results in the metallicity, depending on the index used. Fig. \ref{ionization} shows the ionization rate of oxygen in the galaxy using the parameter U defined in \cite{Diaz00} as U$=$ --0.80 log([OII] 3726,29/[OIII] 5007) -- 3.02. Due to the offset in the flux calibration between the LR blue and the HR orange grism (see Sect. \ref{observations}), we use the fluxes for both lines measured in the LR blue grism

The variation in the ionization parameter is rather small across the galaxy both in the 2D datacube and the integrated regions. The star-forming generally show a somewhat higher ionization than the bulge, but nowhere [OIII] is particularly strong in comparison to [OII]. Ionization has been shown to correlate with metallicity, with ionization being higher at low metallicities \citep{Dopita06}. The host of GRB 060505 follows more or less this trend, the places of lowest metallicity (GRB, HII-SW and HII-E) also have the highest ionization, although e.g. the ionization of HII-S with one of the highest metallicities is very similar to the one of the GRB site. Taking the ionization into account would imply metallicities of around 0.1 dex higher for the regions of higher ionization (see e.g. \cite{KD02}) but no correction for the regions of lower ionization such as the bulge. We leave the values here without any correction for ionization.

\subsection{Star formation and age}

To determine the star-formation rate we use the conversion between the H$\alpha$ emission line and the SFR as derived by \cite{Kennicutt}, $\mathrm{SFR}$[M$_\odot$/yr]=$7.9\times10^{-42} 4 \pi f_{H\alpha} d_L^2$ with d$_L$ being the luminosity distance. The H$\alpha$ emission is directly linked to the UV flux by young, massive stars and suffers relatively little from a possible extinction in the galaxy. The host has a total SFR of 1.2 M$_\odot$/yr, similar to the MW, but relatively high considering the low luminosity of $\sim$0.1 L* of the galaxy.

An estimate for the actual ongoing star-formation rate compared to the luminosity of the region is obtained by weighting the SFR with the B-band luminosity. We determine the B band magnitude by summing the flux in the B-band range from the LR blue grism data. The fluxes are converted to absolute magnitudes and we then multiply the SFR with the ratio between the absolute magnitudes and the magnitude of an L* galaxy M$_\mathrm{B}=$-21\,mag in order to obtain the specific (luminosity-weighted) star-formation rate (SSFR) using the following conversion: $\mathrm{SSFR}$=$\mathrm{SFR}\times10^{0.4(M_B+21.)}$. The values for the integrated spectra are then normalized to the value per spaxel to account for the different sizes of the extracted regions.

While the SSFR generally indicates a somewhat elevated SF throughout the galaxy, the SSFR increases from the centre to the spiral arms of the galaxy as it is expected for an inside-out starformation scenario in spiral galaxies. The highest SSFR is found in the SW-spiral arm that is also very prominent in the H$\alpha$ image of the galaxy while the rest of the HII regions show similar values for the SSFR. The SSFR follows the same trend as found in the longslit data of \cite{Thoene08} albeit it is less pronounced as found in the previous work.

\begin{figure}
\includegraphics[width=\columnwidth, angle=0]{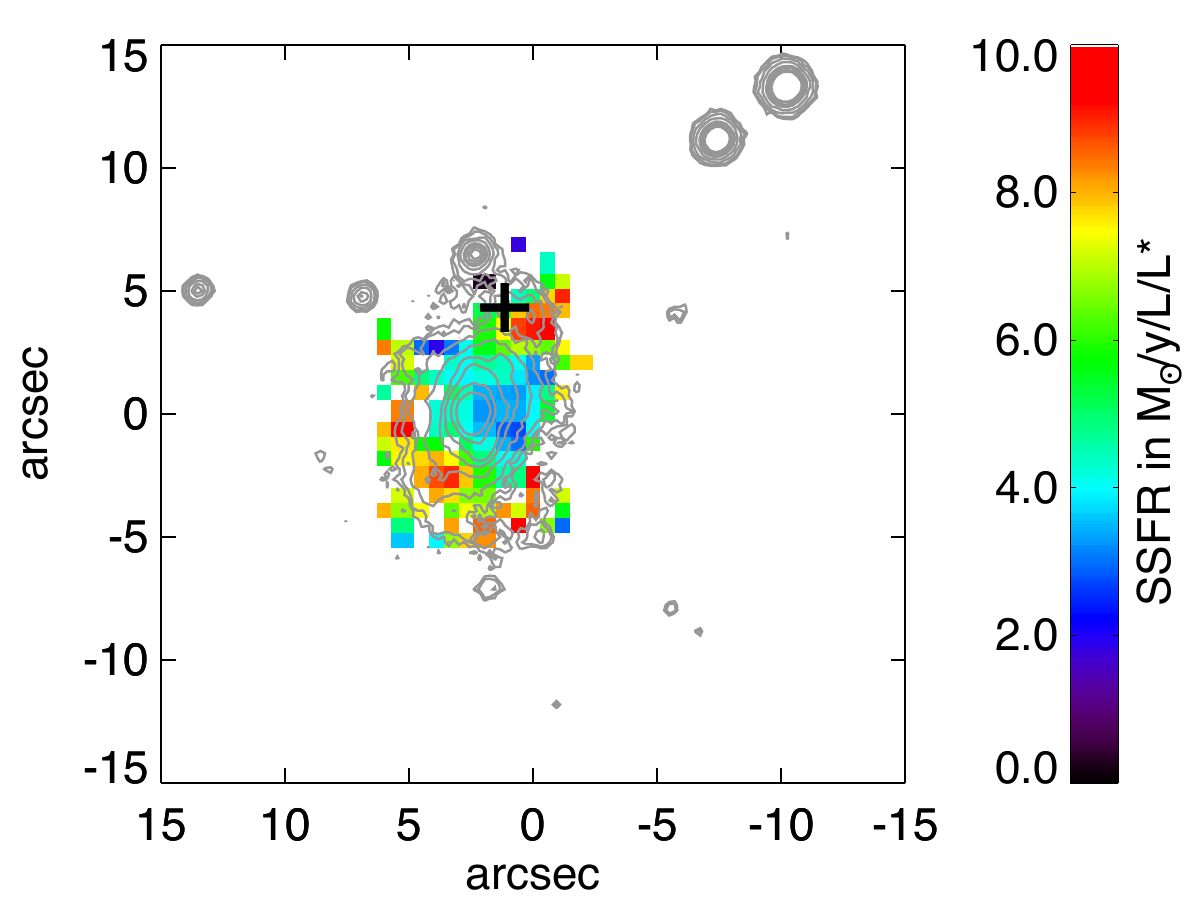}
\caption{SFR weighted by the restframe B-band luminosity in M$_\odot$/yr/kpc/L/L*, the cross indicates the region of the GRB.
\label{SFRabs}}
\end{figure}

As a rough proxy for the age of the stellar population, we use the equivalent width (EW) of H$\alpha$. We can only do this for the integrated regions since the continuum is generally too weak in the individual spaxels. Surprisingly, the EW is not high in all of the star-forming regions. The GRB site only has the third highest EW after the HII-N next to the GRB site and HII-W, which has otherwise no extreme properties. In \cite{Thoene08} an EW of almost -200\AA{} was found (corresponding to an age of $\sim$6Myr), considerably higher than in the other parts of the galaxy along the slit. This difference to the longslit data might be explained by integrating over larger regions comprising parts with lower EWs than it was done for the longslit data, which had been obtained under good seeing conditions of 0.75 arcsec \citep{Thoene08}.

\subsection{Extinction}
We determined the extinction in each spaxel by using the Balmer line decrement  H$\alpha$/H$\beta$ which has a fixed value of 2.76 for case B recombination and zero extinction \citep{Osterbrock}. Regions with a H$\beta$ flux of less than 0.1$\times$10$^{-16}$ erg\,cm$^{2}$\,s$^{-1}$ were excluded from the extinction measurements.

The extinction is low or consistent with zero throughout most regions of the galaxy including the GRB site. Only the bulge shows some extinction, which primarily comes from some more dusty region in the southern part of the bulge. It has already been noted in \cite{Thoene08} that the low extinction excludes that a possible SN to GRB 060505 was hidden by dust. It also implies that the SFR we see is not severely affected by dust and can be taken as close to the absolute SFR in the galaxy. Due to the low extinction in most parts of the galaxy, we do not correct the emission line fluxes for extinction.

\begin{figure}
\includegraphics[width=\columnwidth, angle=0]{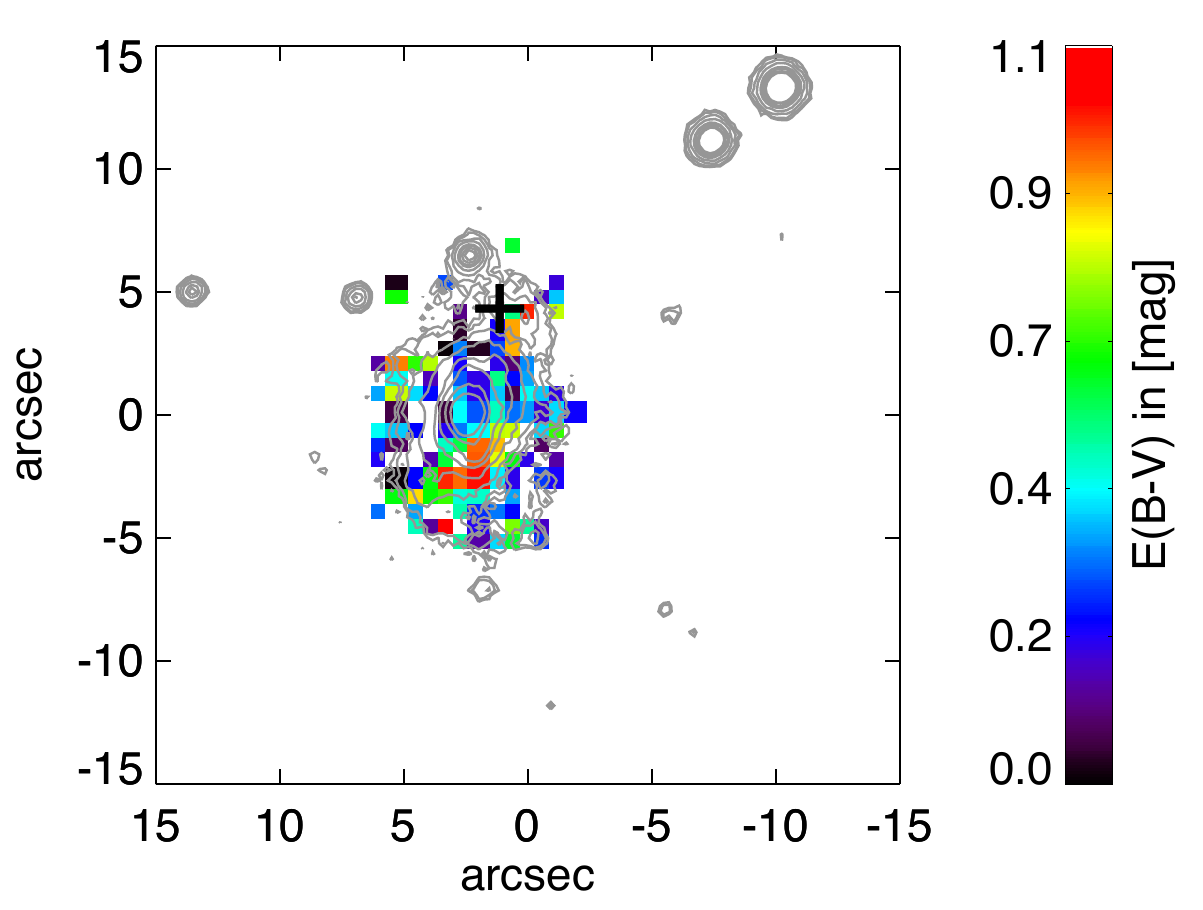}
\caption{Extinction values in E(B--V) using the Balmer decrement of H$\alpha$ vs. H$\beta$, regions with H$\alpha$ $<$ 0.15$\times$10$^{16}$erg\,cm$^{-2}$\,s$^{-1}$ were excluded.
\label{ext}}
\end{figure}

\subsection{Comparison between different HII regions}\label{comparison}
One of the main intentions of the paper is to investigate whether the GRB region had unique properties or whether it was a mere coincidence that the GRB occurred exactly in this SF region. Furthermore, we want to study if the global properties of the galaxy are largely different from the properties of the GRB site as we would have measured it for the same galaxy at a higher redshift. This in turn would then give wrong conclusions for similar cases where the host galaxy had been at higher redshift and unresolved.

\begin{figure}
\includegraphics[width=\columnwidth, angle=0]{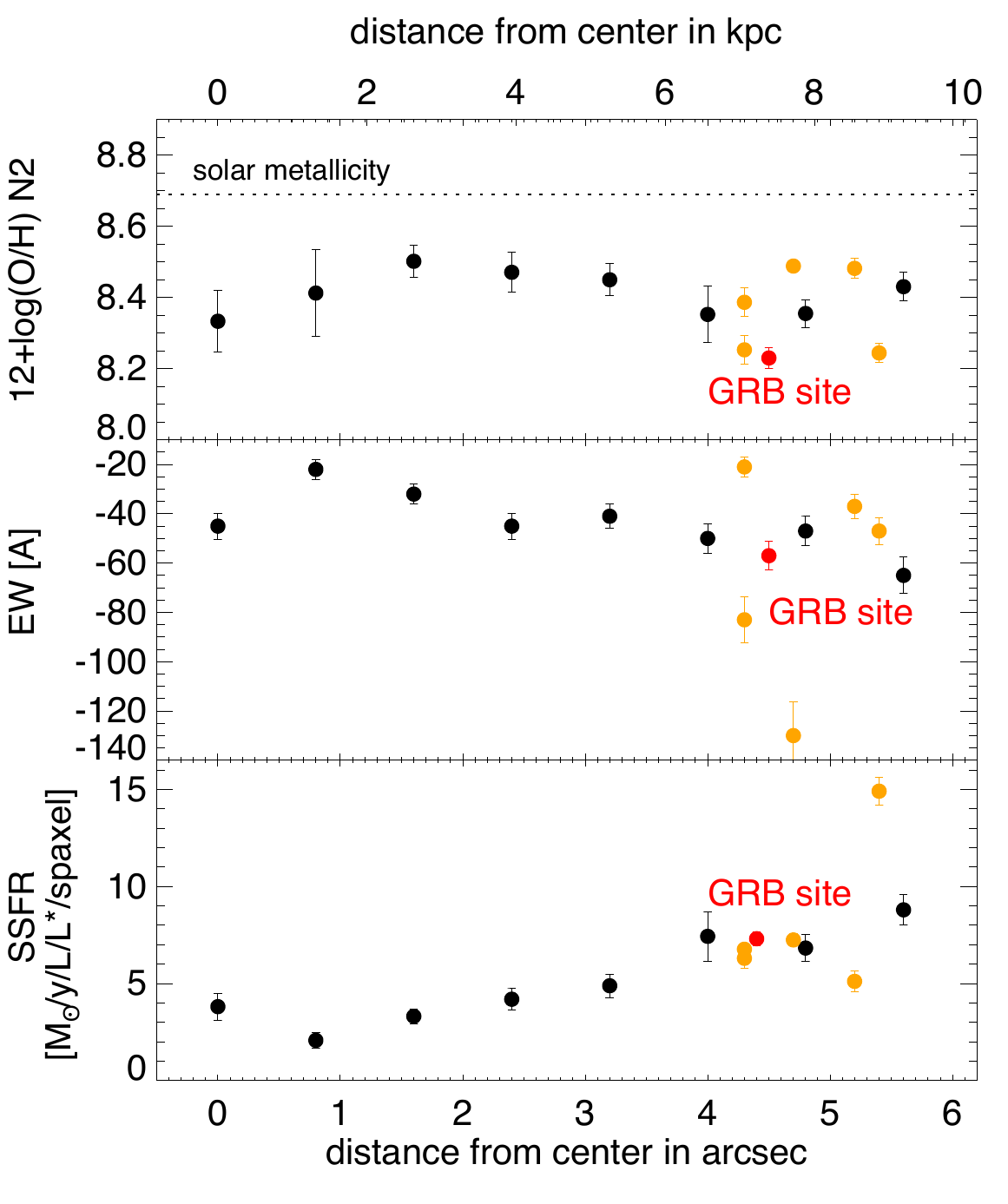}
\caption{Comparison of different properties between the integrated HII regions as show in Fig. 1 (orange points) and the average properties extracted from elliptical regions around the center of the galaxy (black points). The HII regions are plotted at their corresponding deprojected distance from the galaxy centre. The integrated spectrum of the bulge was left out in the plot. For the EW we used the H$\alpha$ line.
\label{ellipses_prop}}
\end{figure}

We compare the different HII regions with the average properties at their corresponding distance from the centre. To do this, we extract integrated spectra of ellipses around the centre in steps of 0\farcs8 using a PA of 38 deg and an inclination of 49 deg (see \cite{Thoene08})\footnote{using the PINGSoft package \citep{Pingsoft} written in IDL: http://centeotl.inaoep.mx/$\sim$frosales/pings/html/software/ }. The properties from these elliptical regions are determined in the same way as for the integrated regions and the results for both the ellipses and the HII regions at their corresponding deprojected distance are plotted in Fig. \ref{ellipses_prop}. Fluxes and derived values for the ellipses are found in Tab.\ref{ellipses:fluxtable} and Tab.\ref{ellipses:comptable} in the appendix. 

The metallicity shows an initial rise and then follows a slight gradient towards the outskirts. The inverse is the case with the EW, indicating a somewhat younger population in the outskirts. The SSFR/spaxel is low in the bulge and rises towards the outskirts. The SSFR of all but one HII region, including the GRB site, is around the average at that distance from the centre, not surprising, since it is exactly those HII regions that contribute to the SFR in the spiral arms. The GRB site is only somewhat different from the average value at its distance from the centre in the metallicity and the SSFR while it has an ``average'' EW and hence age. Except for the EW, the GRB region has the most extreme properties in the galaxy but is not a stark outlier as suggested in the longslit data in \cite{Thoene08}. The other HII regions are rather scattered in their properties, but all have rather high SSFRs compared to their position in the galaxy, not surprising since they were defined as such by their large H$\alpha$ fluxes.

\begin{figure}
\includegraphics[width=\columnwidth, angle=0]{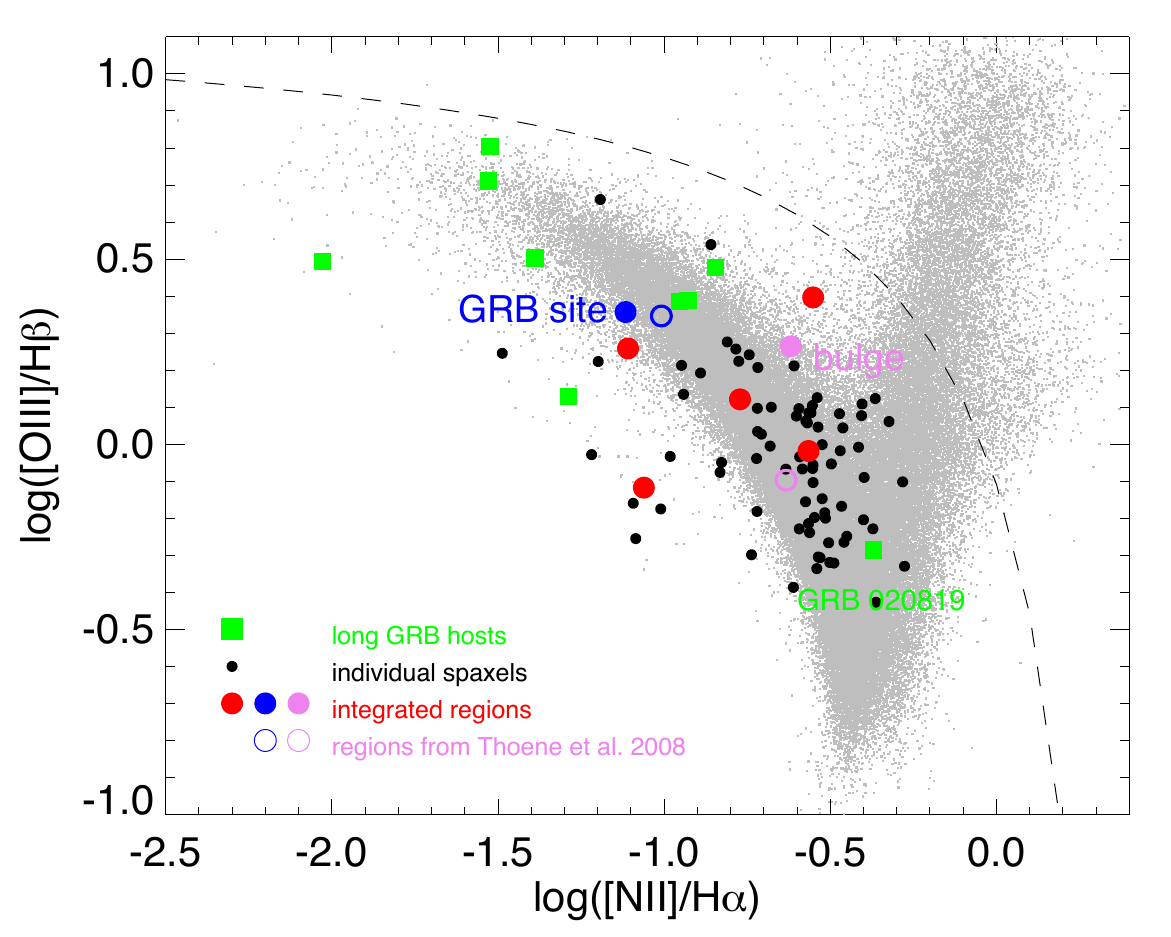}
\caption{BPT diagram of NII/H$\alpha$. In grey are emission line galaxies from the SDSS DR 10, green are other z$<$0.5 long GRB hosts, most of them are data from unresolved hosts. Black dots represent individual, filled circles integrated regions in the host of GRB 060505 from the VIMOS data. The dashed line is the dividing line between AGN and HII regions as derived in \citealt{Kewley06}. For comparison, we also show the line ratios of the GRB site and the bulge as derived from the longslit data on the host from \citealt{Thoene08} (blue and purple empty circles). 
\label{plot:BPT}}
\end{figure}

We put the values from the integrated regions as well as individual spaxels in the ``Baldwin, Philips and Terlevich" diagram (Fig.\ref{plot:BPT}) where we plot [NII]/H$\alpha$ vs. [OIII]/H$\beta$, which are also used for determining metallicities (see Sect. \ref{sect:metal}). This diagram allows to distinguish between region excited by radiation from young stars and such excited by AGNs. All regions in our host are clearly not affected by any AGN activity. The diagram furthermore gives some indications of metallicity and age with metallicity decreasing towards the left (since [NII]/H$\alpha$ is a metallicity calibrator) and the age increasing from top to bottom. The GRB site is hence one of the more metal poor and younger regions in the host but not as extreme as a number of other long duration GRB hosts.

\begin{figure*}
\includegraphics[height=5.2cm, angle=0]{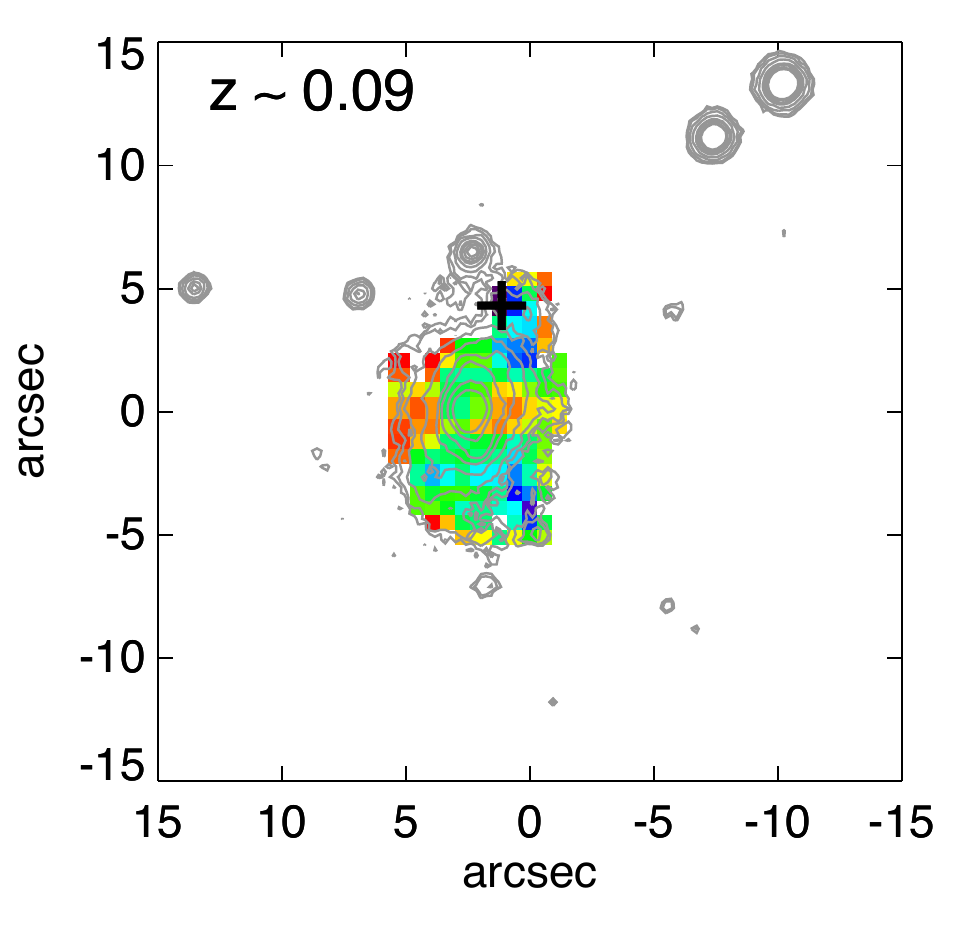}
\includegraphics[height=5.2cm, angle=0]{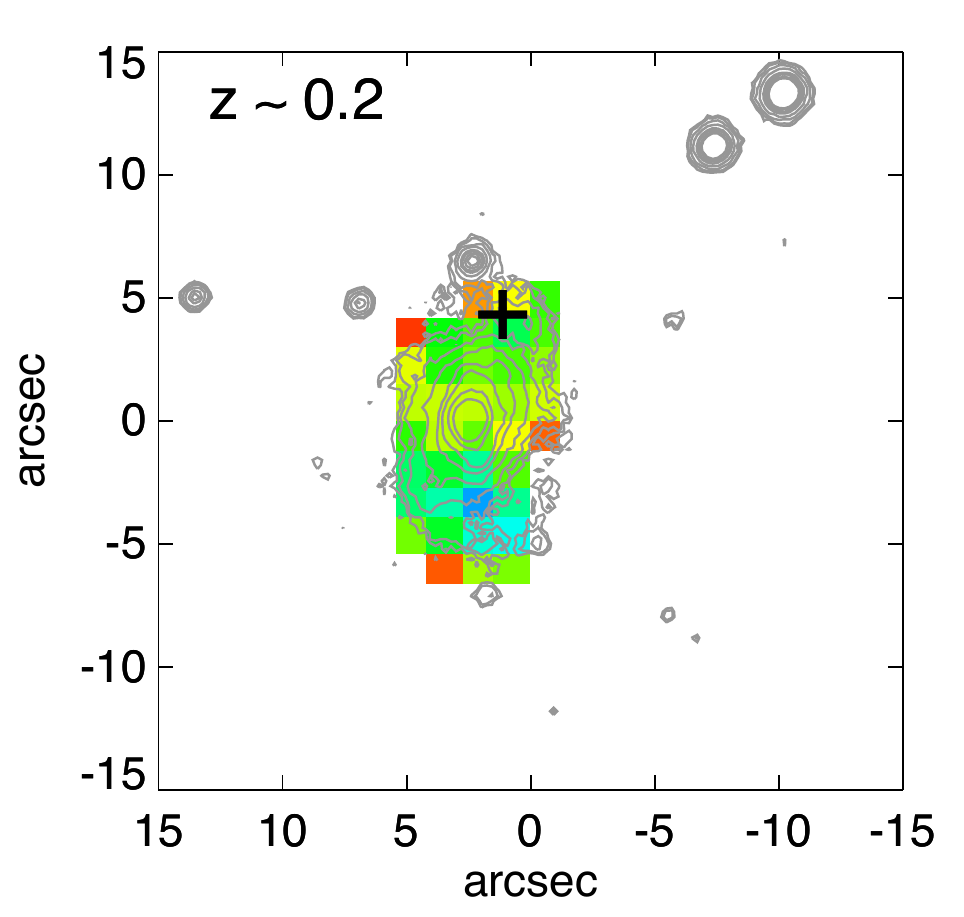}
\includegraphics[height=5.2cm, angle=0]{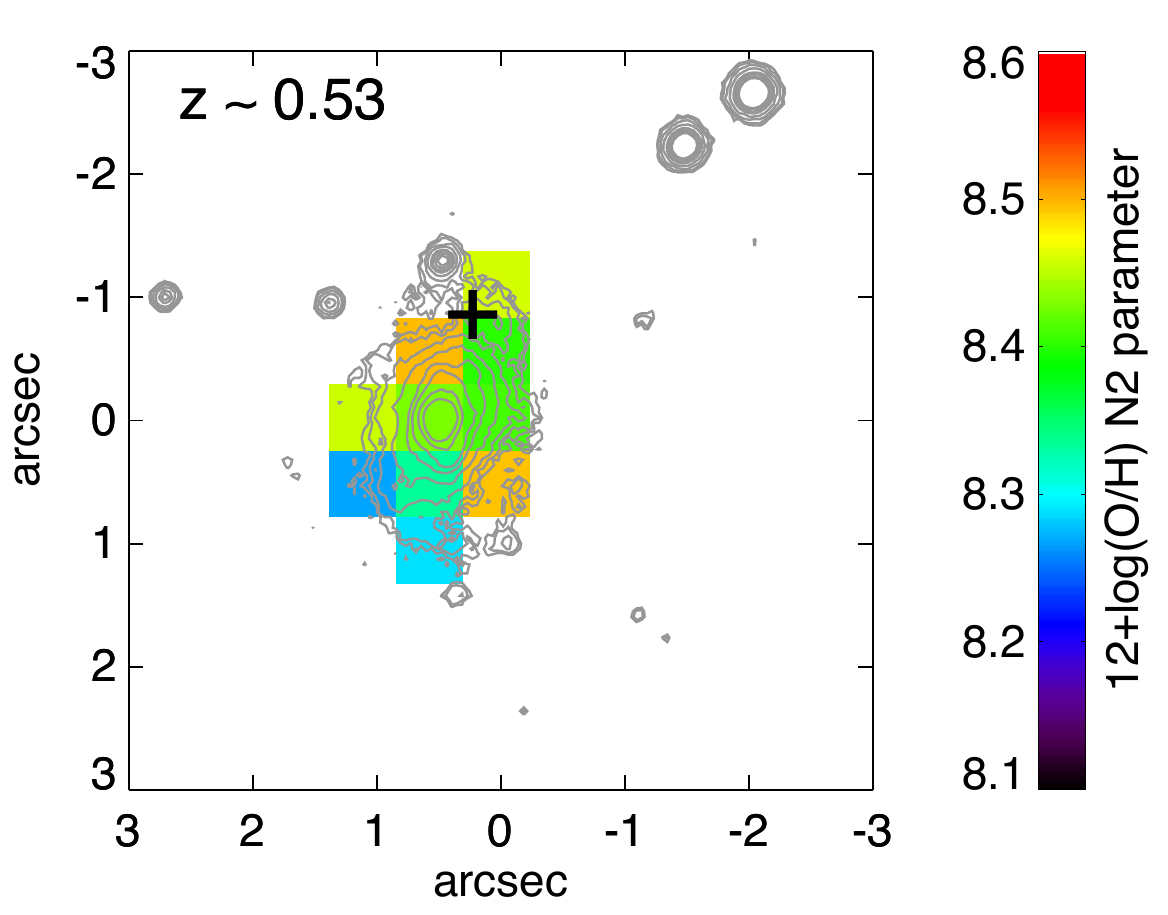}
\caption{Simulation of the metallicity (N2 index) distribution in the host if the galaxy had been at redshifts of z$\sim$0.2 and 0.53, in comparison to the actual position at z$=$0.09. The color scale varies slightly between the different plots for better visualization of the metallicity scale. The data have only been rebinned, lower S/N due to the larger distance was not taken into account.}
\label{downgrade}
\end{figure*}

An important issue when making any conclusions on the progenitor from measurements of the GRBs environment or host is the spatial resolution. Depending on the size of the probed region and how properties vary across the host determine how accurately the measurement at the ``GRB site'' reflect the actual environment. In order to test this, we rebin the datacube of the GRB 060505 host to resolutions the VIMOS data would have given us if the host had been at z$=$ 0.2 and 0.53 and compare it with the values of the original datacube (see Fig. \ref{downgrade}). Any decrease in resolution generally wipes out locally confined more extreme properties. The metallicity gradient becomes much shallower and any difference between the HII regions around the centre basically disappears.

The conclusions derived from longslit data on the other hand crucially depend on the placement of the slit. In our case the GRB region was sticking out in its properties compared to the other parts probed along the slit and also compared to the HII region at the opposite arm. This is a chance coincidence since we probed the tip of the HII-S region which actually has the highest metallicity among the HII regions. Regions similar to the GRB site like the bright HII-SW region were completely missed as would have been any HII region outside the slit with properties very different from the GRB site. By placing a longslit through the centre of a barely resolved large host at high redshift could have a similar effect since a significant fraction of the HII regions in the spiral arms would be missed in the spectrum and the analysis. 

\subsection{Analysis of the HIRES data}
The HIRES data allow us to study the metallicity, extinction and [SII]/H$\alpha$ along slit 1 which goes from the centre of the galaxy to the GRB site. For slit 2, [NII] and [SII] are too weak to be fitted and we can therefore only derive the ionization and extinction but no metallicity.

Concerning the metallicity, the HIRES data give consistent values for the N2 and the O3N2 index. We can only determine the metallicity just before the GRB site since [NII] is not detected with high enough significance at the GRB site whereas H$\alpha$ is clearly visible, which further points to a low metallicity. The metallicity just next to the GRB region is 0.1 dex higher than the values found in the integrated spectrum of the GRB site from the VIMOS data. The metallicity gradient shows the same behaviour as for the VIMOS integrated spectra at different distances from the centre reaching the highest value at the outskirts of the bulge, not in the centre, and subsequently falling towards the outskirts.

Due to the better resolution allowing a better skyline subtraction, we can determine the ratio log ([SII]$\lambda$ 6717,6731/H$\alpha$) along slit 1 in the HIRES data. [SII]$\lambda$ 6731 has too low S/N for a proper extraction and the ratio of the two SII lines can generally vary between $\sim$ 1.5 for low electron density and $\sim$0.5 for high densities. From the longslit data we know that [SII] 6717/ [SII] 6731 is always $>$1.25 and we therefore take a ratio of 1.3 for all the regions, hence the values for [SII]/H$\alpha$ are hence considered a lower limit. log([SII]/H$\alpha$) has values similar to what has been found in nearby spiral galaxies and there is no indication for regions excited by shocks which would give ratios of [SII]/H$\alpha$ around 1. The extinction along both slits is very low or not measurable basically along all the slit, consistent with what we found for the VIMOS data.

\begin{figure}
\centering
\includegraphics[width=\columnwidth]{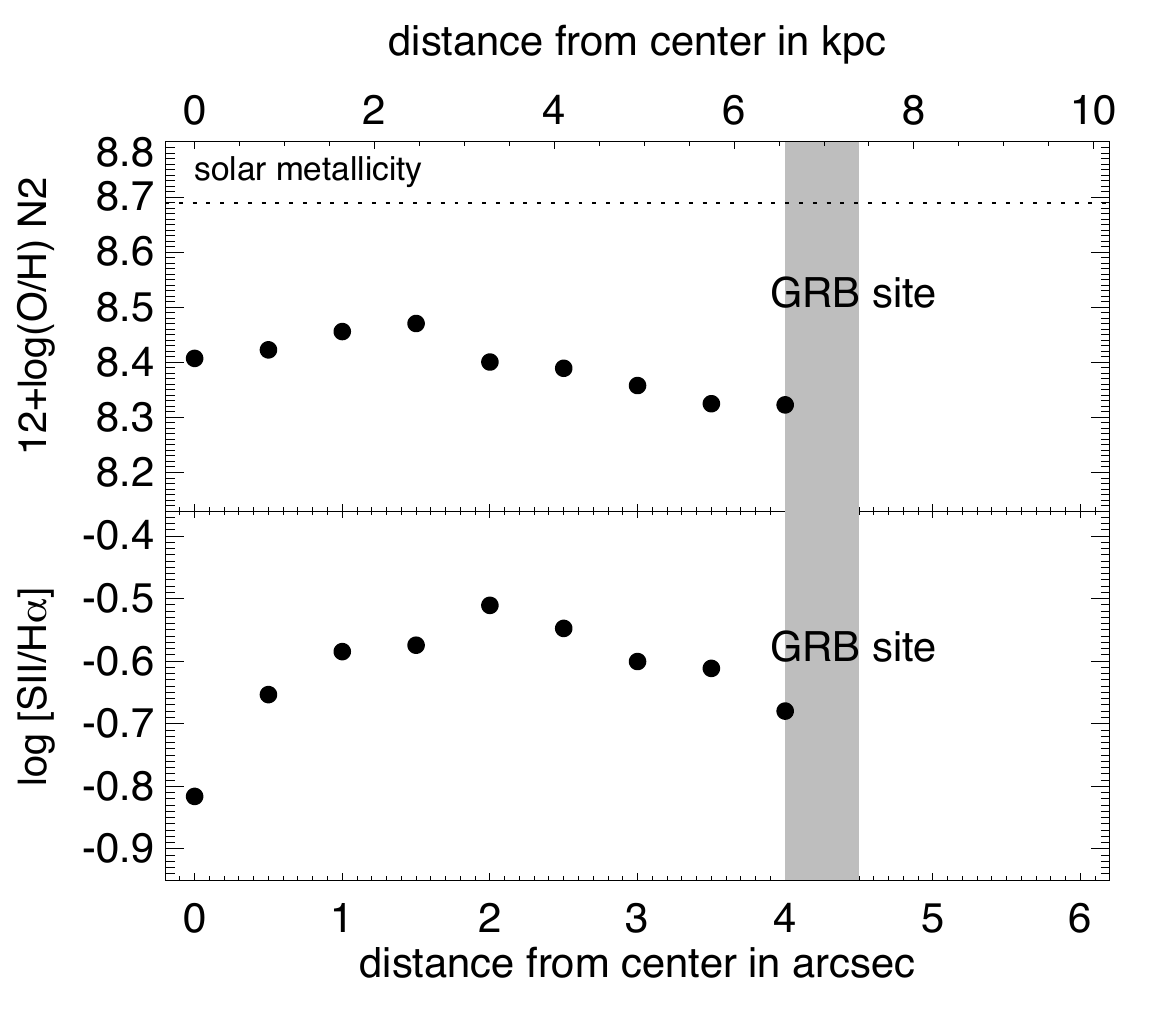}
\caption{Metallicity and [SII]/H$\alpha$ along slit 1 of the HIRES data. The values are listed in Tab.\ref{table:HIRESval} in the appendix. \label{HIRESprop}}
\end{figure}

\section{Kinematics of the galaxy}

The host of GRB 060505 is an isolated dwarf spiral with no obvious signs of interaction. The large-scale environment confirms that it does not belong to a larger group, but it might be part of a filament extending from the Abell 3837 cluster \citep{Thoene08}. Interaction has been one possible cause for an increase in star-formation which again facilitates the formation of very massive stars and hence GRBs, but SF can also be triggered by gas inflow or compression due to supernova explosions. In the following, we investigate the resolved kinematics of our galaxy to check for external causes for the recent star-formation.

\subsection{Rotation curve fitting}
\begin{figure}
\centering
\includegraphics[width=\columnwidth]{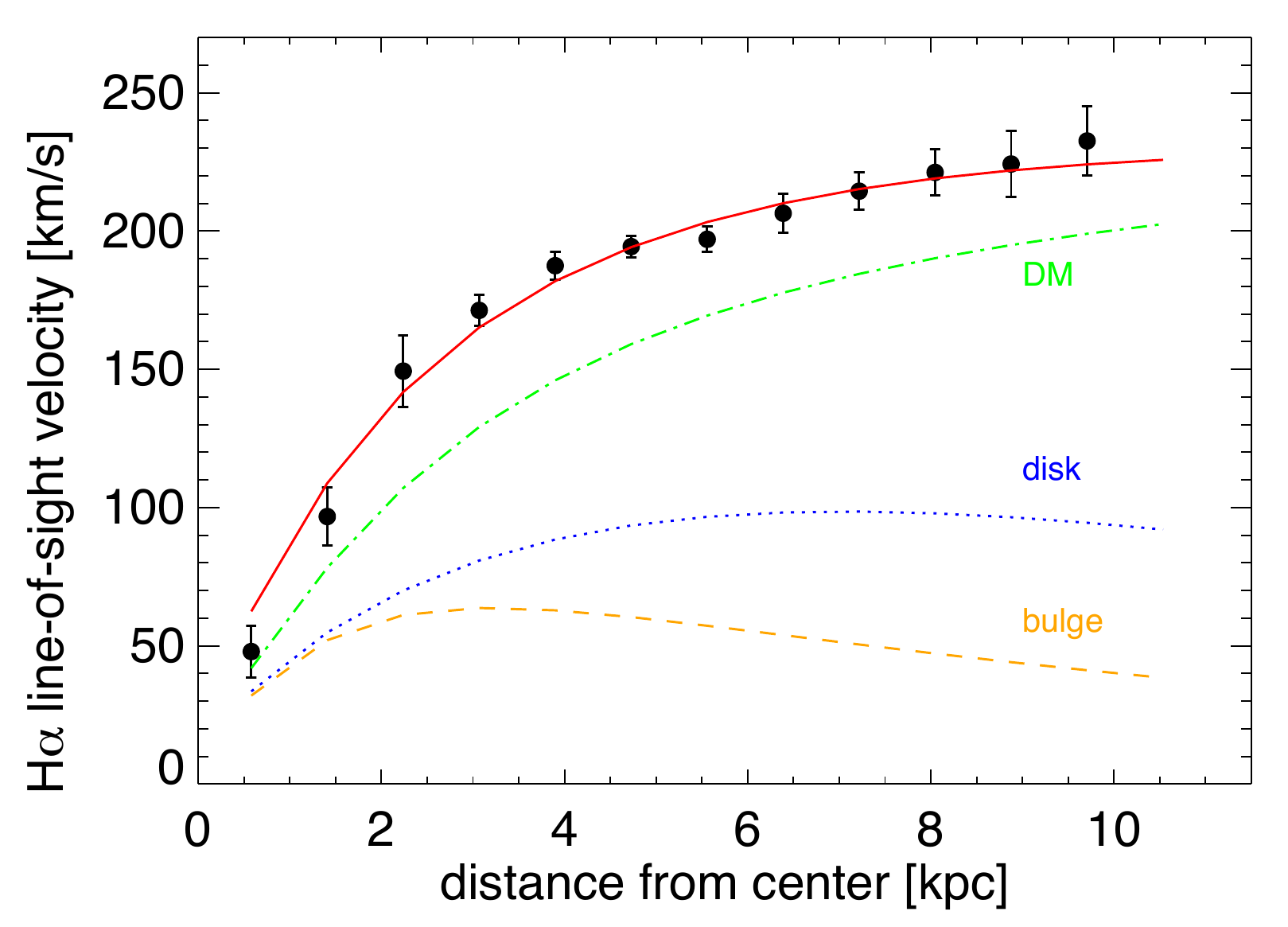}
\caption{Rotation curve for one half of the GRB 060505 host using the H$\alpha$ line of slit 1 of the HIRES spectra corrected for the inclination of 49deg. We fit the rotation curve with disk, bulge and a DM halo component, as DM halo profile we used an isothermal sphere. The velocities of H$\alpha$ are listed in Tab.\ref{table:HIRESval} in the appendix.}\label{060505:revisedrotcurve}
\end{figure}

In \cite{Thoene08} we derived a rotation curve using H$\alpha$ from low-resolution longslit data. Here we improve the analysis by using the high-resolution longslit data from the HIRES spectrograph which cover the region roughly from the centre or the galaxy to just beyond the GRB region. We take the strongest emission line, H$\alpha$ and extract the line profiles for each bin along the spatial direction where one bin corresponds to one row in the spectrum which is 0.5 arcsec or 0.8 kpc. These profiles are then fit by Gaussians, the centers of which are plotted in Fig. \ref{060505:revisedrotcurve}, corrected for the inclination of 49 deg \citep{Thoene08}.

We then fit this rotation curve with contributions from the disk, the bulge and a dark matter (DM) halo where we take the DM-profile of an isothermal sphere. Other DM profiles would only be different in the outermost regions which we cannot probe with the strong emission lines from SF regions. We get a best fit with the following values ($\chi^2$ = 9.4 with 6 d.o.f): a disk scale length of 8.2$\pm$1.6 kpc, a mass-to-light ratio of 6.8$\pm$1.3 (solar units), a central luminosity density of 59.8$\pm$11.4$\times$10$^{-3}$M$_\odot$/pc$^{3}$, a DM core radius of 2.8$\pm$0.3 kpc and a central density of the DM halo of 144$\pm$35$\times$10$^{-3}$ M$_\odot$/pc$^{3}$ and a bulge radius of 2.3$\pm$0.3 kpc.

The rotation curve does not show any obvious particularities, only the fit in the innermost few kpc is not very good as the rotation curve rises steeper than expected. A steep rise has been observed in a number of spiral galaxies, especially when they have massive bulges \citep[e.g.][and references therein]{Sofue01}. The data quality, however, is too poor make any firm conclusion on a possible additional component or a high mass concentration in the center.

\subsection{Velocity map and dispersion}
\begin{figure}
\centering
\includegraphics[width=\columnwidth]{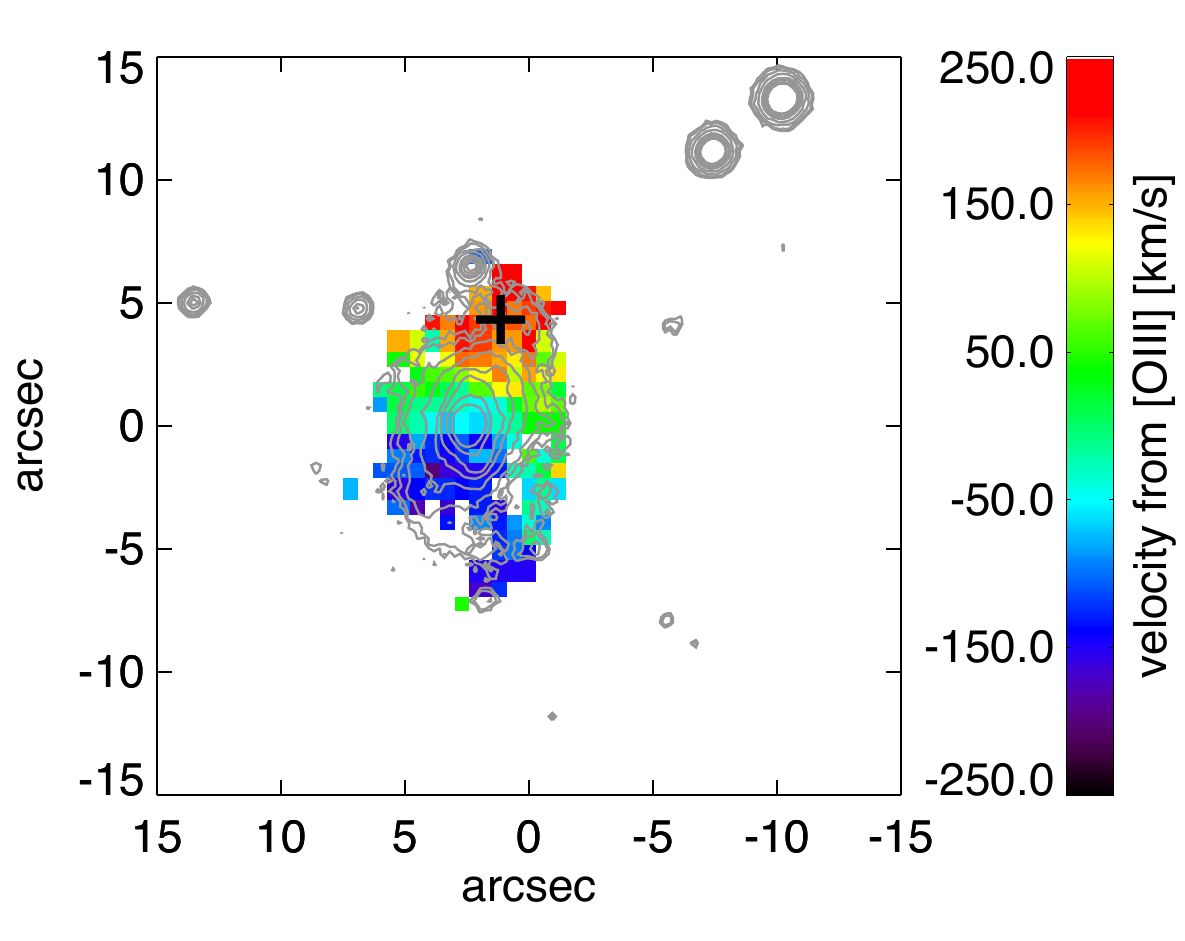}
\caption{Velocity map determined from the center and FWHM of the [OIII] emission line in the VIMOS data.}\label{060505:velcurve}
\end{figure}

\begin{figure}
\centering
\includegraphics[width=\columnwidth]{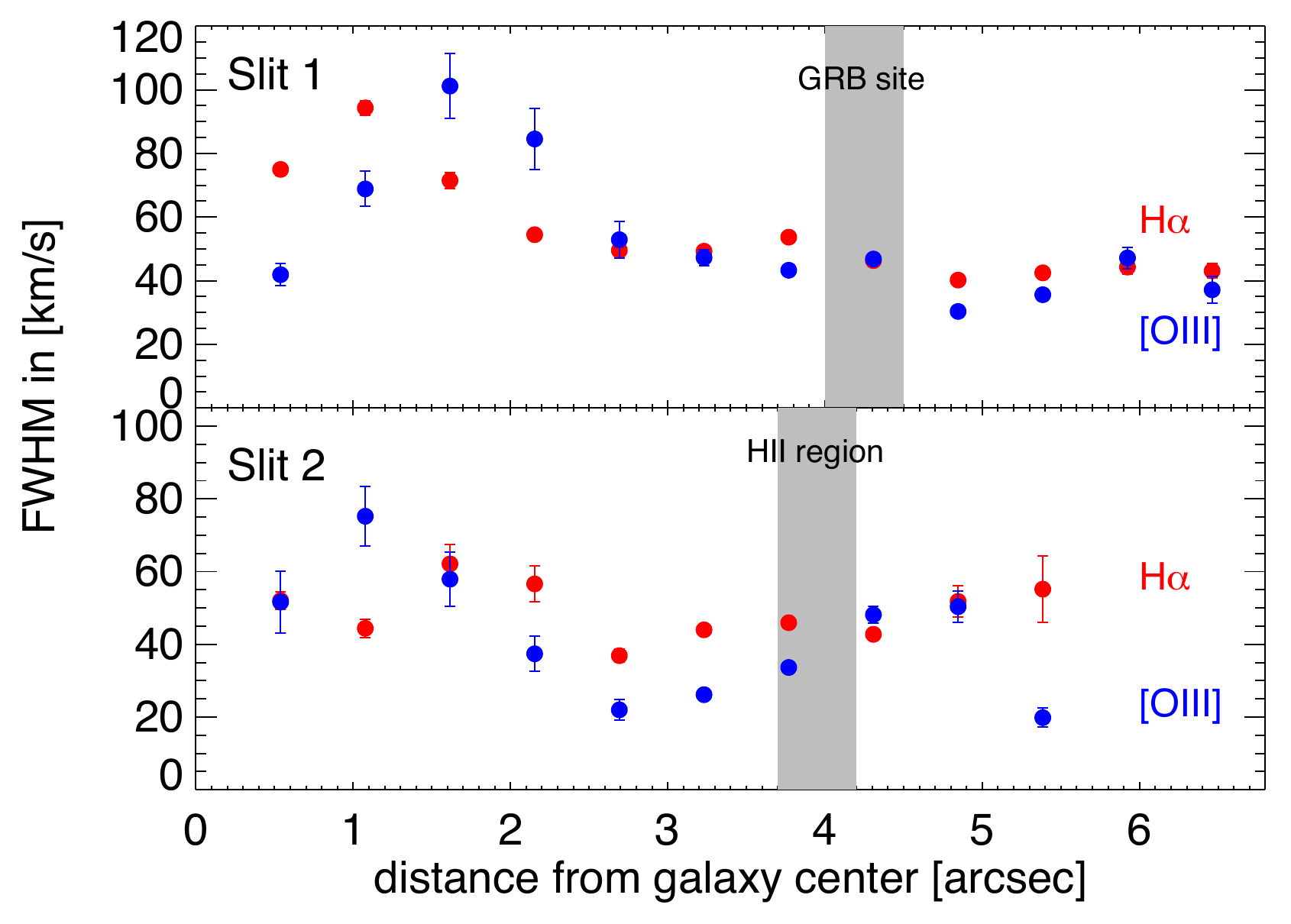}
\caption{Velocity dispersion along the two slits for H$\alpha$ and [OIII], slit 1 includes the GRB region, slit 2 the HII regions next to it (HII-N following the naming in the VIMOS data). The shaded lines indicate the position of the GRB region and the HII region next to it respectively. The values are listed in Tab.\ref{table:HIRESval} in the appendix.}\label{060505:FWHMplot}
\end{figure}

We plot the 2D velocity map from the VIMOS data by fitting the [OIII] emission line with a gaussian fit for all spaxels where the line is detected with large enough significance. [OIII] was used instead of the somewhat stronger H$\alpha$ line due to problems in the shape of H$\alpha$ in a few spaxels in the center of the galaxy, which is probably due to bad pixels. In Fig. \ref{060505:velcurve}, we plot the velocity of the line center taking the center of the galaxy (z=0.0901) as 0 km\,s$^{-1}$. We further analyze the line width in the radial direction from the two slits of the HIRES data using the strongest lines H$\alpha$ and [OIII] plotted in Fig.\ref{060505:FWHMplot}.

The velocity field is rather homogeneous at first glance showing a regular rotating disk. We will investigate this further in the next chapter. The FWHM of the emission lines is around 100 km\,s$^{-1}$ in the centre and decreases outwards towards the spiral arms as seen in the radial HIRES slit (slit 1). An enhancement in the central line width is an indication for gas infall forming a bulge whereas very young disk galaxies show essentially a flat velocity dispersion along the axis \citep[for a small overview over spiral galaxy rotation curves and velocity dispersions see][]{Beckman04}. The line width falls rapidly and remains constant at about the same distance from the center where the rotation curve starts to flatten which is also what we observe here. The difference in velocity between the centre and the spiral arms suggests that the bulge is not very large as is the case in more evolved spiral galaxies.

\subsection{Kinemetry analysis}

\begin{figure}
\centering
~~~\includegraphics[width=6cm]{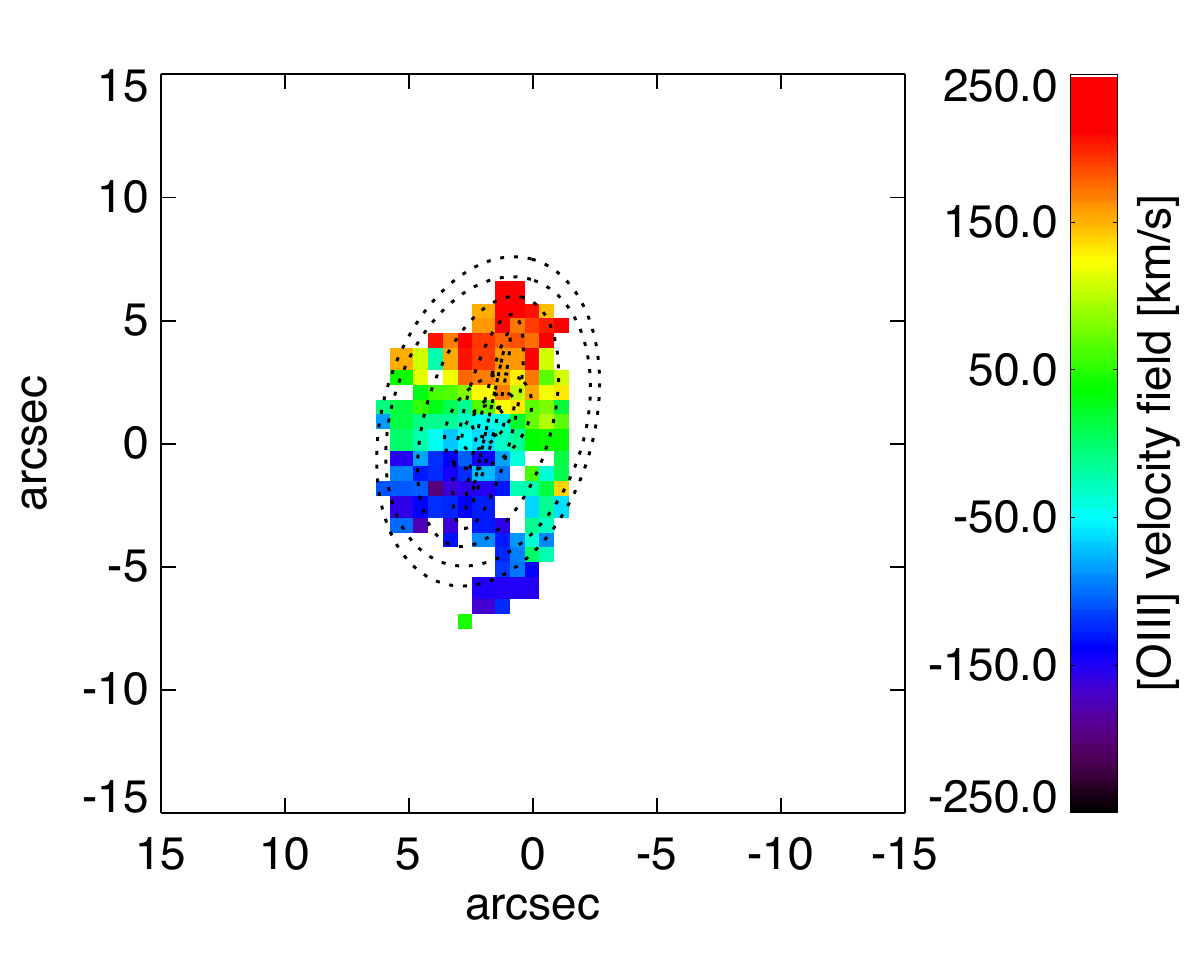}\\
\includegraphics[width=\columnwidth]{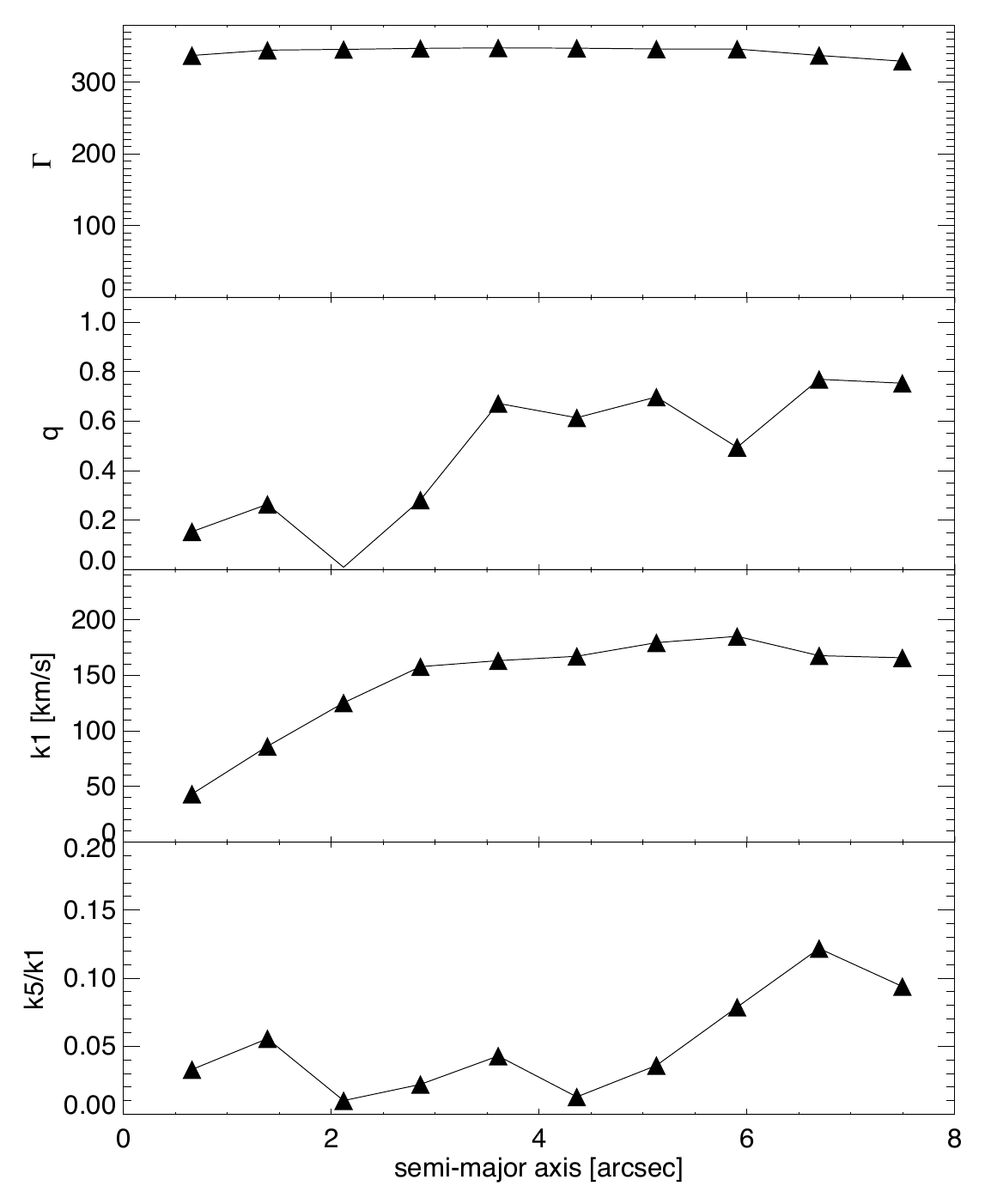}
\caption{Top part: Best fit kinemetry ellipses overplotted on the velocity map derived from [OIII]. Bottom part: Best fit parameters at the different positions relative to the centre of the galaxy. $\Gamma$ is the PA, q the flattening parameter or ellipticity, k1 is the dominant coefficient of odd kinematic moments and indicates the amplitude of the bulk motion for velocity maps. k5/k1 is a measure for the deviation from simple rotation since k5 is the first higher-order term only present in case of perturbations from normal rotation.}\label{kinemetry}
\end{figure}

We further analyze the kinematics of the galaxy by fitting the velocity map with kinemetry models as detailed in \cite{Kinemetry}\footnote{The IDL fitting code is available at the following address: http://www-astro.physics.ox.ac.uk/$\sim$dxk/idl/}. The method derives kinematical information by fitting ellipses to the velocity map described by a harmonic expansion. Higher order moments of the expansion give an indication of perturbed velocity fields. The best fit ellipses overplotted on the velocity map as well values for the PA, ellipticity and the coefficients k1 and k5 of the harmonic expansion are shown in Fig.\ref{kinemetry}.

Our galaxy overall has an axisymmetric velocity field, the PA of the best fit ellipses remains constant across the galaxy and k1 is identical with the rotation curve of the galaxy. The flattening or ellipticity parameter q changes its value at around 4 arcsec which could be indicative of a separate component in the inner parts of the galaxy, but the data are not good enough for a definite conclusion. The jump in q in the third fitted ellipses is due to an anomaly in the data and likely not real. The ratio k5/k1 is indicative of deviations from simple rotation in the map and overall rather low but still not negligible. The higher value in the outskirts could point to some perturbations that created the large SF region such as the one to the SW, but at this point we cannot exclude that these deviations are purely a result of the relatively low data quality for this kind of analysis.

Kinemetry has mostly been used for nearby galaxies, although comparisons with model galaxies allowed to distinguish between mergers and unperturbed systems at z$\sim$2 \cite{Shapiro08} with even lower spatial resolution and data quality. According to the classification in this paper using the asymmetry in the velocity and dispersion map derived from the coefficients of the harmonic expansion v$_{asym}$ vs $\sigma_{asym}$ our galaxy lies in the transition between unperturbed galaxies and mergers. This could be an indication that the SF in this galaxy might be somehow related to some merger event in the past.

\section{Discussion and conclusion}
The host of GRB 060505 was only the second GRB host that has been studied with IFU spectroscopy. It had also the second lowest redshift ever measured for a GRB, making it a fortunate case to study the host in detail, which is currently not possible for the majority of GRB hosts. Studies like the one presented here are crucial to determine the validity of future resolved host studies at higher redshift. GRB hosts at low redshift have proven to be a very mixed bag of galaxies, probably because those bursts would go undetected at higher redshifts. This makes it even more important to carefully study those hosts in detail and to investigate possible biases.

The earlier study with a longslit across the major axis of the galaxy presented in \cite{Thoene08} suggested that the GRB site is a region with rather peculiar properties compared to the rest of the galaxy. Our new study shows that most of the star-forming regions in that galaxy around the bulge in various spiral arms have similar properties to the GRB site and the strong conclusion was mainly an issue of the longslit placement. The GRB site, however, does have the lowest metallicity (albeit within errors) in the host galaxy. 

The kinematic analysis shows a rather smooth velocity field and rotation curve. The high dispersion in the centre might point to some gas inflow to an existing yet not very evolved bulge. A thorough kinemetric analysis reveals some degree of disturbance in the velocity field and we cannot exclude some minor merger event in the past. The host overall has no really extreme properties and the SF in the region around the GRB could simply be due to the inside-out SF propagation usually observed in spiral galaxies. 

Our analysis shows the importance to study low redshift GRB hosts with spatially resolved techniques to be able to properly interpret the global spectra of the entire galaxy that we have to deal with for the large majority of GRB hosts. GRB host are mostly dwarf galaxies, nevertheless they can have different properties throughout the host which is even more relevant for larger hosts. Steep metallicity gradients can lead to wrong metallicities at the GRB site since they are usually found in the outskirts of their hosts, f.ex. determining the metallicity of the integrated spectrum in a larger host seen edge on where the GRB is seemingly lies at the brightest spot of the galaxy but in reality is found in some outer spiral arm (in front or even behind the galaxy since GRBs easily outshine their hosts) would lead to a wrong value. 

Integral field studies of galaxies at all redshifts have become an extremely growing field. Several past and ongoing large IFU such as CALIFA \citep{Sanchez12} or VENGA \citep{Blanc13} have revealed a lot of information about the gas properties and star-formation history in different kinds of galaxies,  but so far they target large nearby galaxies different from the host of GRB 060505 and most other GRB hosts. With future large IFU surveys such as MANGA \footnote{http://www.sdss3.org/future/manga.php} we might even be lucky to observe a galaxy before and after it hosted a GRB to determine if and how such a large explosion influences its immediate environment.

\section*{Acknowledgments}
CCT thanks E. P\'erez Jim\'enez for comments on the manuscript and A. de Ugarte Postigo for cross calibrating the VIMOS fluxes with broad-band photometry. The data presented in this paper were obtained at the W.M. Keck Observatory, which is operated as a scientific partnership among the California Institute of Technology, the University of California and the National Aeronautics and Space Administration. The Observatory was made possible by the generous financial support of the W.M. Keck Foundation. We want to thank the observatory staff of Keck and the VLT for excellent support carrying out the observations presented here. CCT and JG are partially supported by YA2012-39362-C02-02. LC is supported by the EU under a Marie Curie Intra-European Fellowship, contract PIEF-GA-2010-274117. The ``DARK Cosmology Centre'' is funded by the Danish National Research Foundation. This study has been developed in the framework of the Unidad Asociada IAA/CSIC-UPV/EHU. JG acknowledges  the support of the Ikerbasque science foundation to the Unidad Asociada IAA/CSIC-UPV/EHU. JPUF acknowledges support from the ERC-StG grant EGGS-278202. PJ acknowledges support by a Project Grant
from the Icelandic Research Fund.

\appendix

\begin{table*}
\scriptsize
\caption{Fluxes and various properties from the VIMOS data. Fluxes are in 10$^{-16}$ erg\,cm$^{-2}$\,s$^{-1}$. }
\begin{tabular}{lllllllllll}\hline\hline
Spaxel & H$\alpha$ & H$\beta$ & [NII] & [OIII] & [OII]& Z (N2) & Z (O3N2) &log U & SSFR & E(B--V) \\
 & & & 6586 & 5007 & 3727,29& 12+log(O/H)&12+log(O/H)& [OII]/[OIII]& M$_\odot$/yr/L/L* & [mag]  \\ \hline\hline
20 11 &    0.15 &   0.02 &   0.00 &   0.08 &   0.68 &   0.00 &   8.44 &  -2.87 &    5.29 &    0.00\\
21 11 &    0.15 &   0.02 &   0.00 &   0.06 &   0.84 &   0.00 &   8.54 &  -3.31 &    7.60 &    0.00\\
19 12 &    0.18 &   0.06 &   0.01 &   0.05 &   0.20 &   8.32 &   8.42 &  -2.78 &    6.94 &    0.00\\
20 12 &    0.27 &   0.06 &   0.00 &   0.07 &   0.57 &   8.03 &   8.30 &  -2.98 &   11.25 &    0.00\\
21 12 &    0.25 &   0.05 &   0.00 &   0.06 &   0.65 &   8.03 &   8.35 &  -3.15 &   11.93 &    0.06\\
22 12 &    0.16 &   0.04 &   0.00 &   0.07 &   0.68 &   0.00 &   8.39 &  -3.24 &   10.67 &    0.00\\
19 13 &    0.22 &   0.07 &   0.02 &   0.07 &   0.19 &   8.23 &   8.43 &  -3.43 &    8.87 &    0.00\\
20 13 &    0.31 &   0.06 &   0.03 &   0.08 &   0.44 &   7.97 &   8.23 &  -3.32 &   12.36 &    0.00\\
21 13 &    0.27 &   0.05 &   0.00 &   0.09 &   0.48 &   0.00 &   8.14 &  -3.32 &   12.40 &    0.08\\
22 13 &    0.18 &   0.04 &   0.00 &   0.10 &   0.66 &   0.00 &   8.24 &  -3.13 &   10.15 &    0.00\\
18 14 &    0.16 &   0.03 &   0.03 &   0.04 &   0.34 &   8.57 &   8.47 &  -2.82 &    7.93 &    0.04\\
19 14 &    0.17 &   0.06 &   0.00 &   0.09 &   0.21 &   0.00 &   8.42 &  -3.31 &    8.42 &    0.05\\
20 14 &    0.26 &   0.06 &   0.00 &   0.12 &   0.36 &   8.14 &   8.24 &  -3.20 &   12.46 &    0.06\\
21 14 &    0.25 &   0.06 &   0.00 &   0.14 &   0.25 &   0.00 &   8.25 &  -3.01 &   11.49 &    0.00\\
22 14 &    0.24 &   0.06 &   0.00 &   0.15 &   0.53 &   0.00 &   8.25 &  -2.79 &   13.19 &    0.00\\
23 14 &    0.16 &   0.07 &   0.00 &   0.14 &   0.53 &   0.00 &   8.45 &  -2.69 &    8.61 &    0.00\\
13 15 &    0.15 &   0.06 &   0.03 &   0.04 &   0.21 &   8.25 &   8.34 &  -3.18 &    7.21 &    0.00\\
17 15 &    0.19 &   0.03 &   0.07 &   0.05 &   0.28 &   8.52 &   8.48 &  -2.94 &    8.30 &    0.18\\
18 15 &    0.23 &   0.05 &   0.04 &   0.10 &   0.40 &   8.48 &   8.43 &  -2.83 &    8.75 &    0.26\\
19 15 &    0.20 &   0.06 &   0.00 &   0.14 &   0.48 &   0.00 &   8.38 &  -3.22 &    8.67 &    0.19\\
20 15 &    0.25 &   0.06 &   0.02 &   0.17 &   0.49 &   7.96 &   8.25 &  -3.16 &   10.45 &    0.25\\
21 15 &    0.23 &   0.05 &   0.00 &   0.16 &   0.47 &   0.00 &   8.18 &  -3.05 &    9.80 &    0.00\\
22 15 &    0.23 &   0.07 &   0.00 &   0.13 &   0.63 &   0.00 &   8.17 &  -2.88 &   10.59 &    0.03\\
23 15 &    0.17 &   0.08 &   0.00 &   0.10 &   0.80 &   0.00 &   8.18 &  -2.82 &    6.76 &    0.00\\
13 16 &    0.17 &   0.03 &   0.02 &   0.03 &   0.24 &   8.25 &   8.35 &  -3.24 &    8.13 &    0.00\\
15 16 &    0.23 &   0.07 &   0.04 &   0.00 &   0.14 &   8.51 &   8.35 &  -2.66 &    7.63 &    0.00\\
16 16 &    0.23 &   0.06 &   0.03 &   0.01 &   0.21 &   8.50 &   8.45 &  -2.78 &    5.58 &    0.33\\
17 16 &    0.29 &   0.06 &   0.02 &   0.06 &   0.42 &   8.47 &   8.49 &  -3.05 &    6.51 &    0.25\\
18 16 &    0.28 &   0.07 &   0.03 &   0.12 &   2.26 &   8.44 &   8.44 &  -3.12 &    7.50 &    0.41\\
19 16 &    0.22 &   0.06 &   0.00 &   0.16 &   2.42 &   0.00 &   8.39 &  -3.14 &    7.05 &    0.31\\
20 16 &    0.21 &   0.06 &   0.00 &   0.19 &   2.50 &   0.00 &   8.25 &  -3.21 &    8.35 &    0.34\\
21 16 &    0.19 &   0.06 &   0.00 &   0.17 &   0.78 &   0.00 &   8.14 &  -3.03 &    7.34 &    0.09\\
22 16 &    0.21 &   0.08 &   0.00 &   0.18 &   0.74 &   0.00 &   8.13 &  -3.00 &    8.73 &    0.17\\
23 16 &    0.22 &   0.07 &   0.00 &   0.11 &   0.72 &   8.28 &   8.18 &  -3.08 &    8.51 &    0.03\\
13 17 &    0.15 &   0.02 &   0.00 &   0.04 &   0.47 &   0.00 &   8.31 &  -3.19 &    6.21 &    0.00\\
14 17 &    0.17 &   0.05 &   0.00 &   0.04 &   0.38 &   0.00 &   8.25 &  -2.97 &    7.33 &    0.00\\
15 17 &    0.29 &   0.07 &   0.06 &   0.03 &   0.40 &   8.37 &   8.29 &  -2.82 &    9.03 &    0.00\\
16 17 &    0.37 &   0.07 &   0.05 &   0.07 &   0.55 &   8.40 &   8.32 &  -2.85 &    8.22 &    0.25\\
17 17 &    0.41 &   0.08 &   0.08 &   0.13 &   0.78 &   8.40 &   8.40 &  -3.05 &    7.84 &    0.24\\
18 17 &    0.38 &   0.09 &   0.08 &   0.20 &   4.17 &   8.40 &   8.42 &  -3.38 &    6.75 &    0.40\\
19 17 &    0.31 &   0.09 &   0.04 &   0.20 &   3.72 &   8.36 &   8.40 &  -3.39 &    6.64 &    0.40\\
20 17 &    0.27 &   0.09 &   0.02 &   0.19 &   3.66 &   8.31 &   8.35 &  -3.48 &    6.41 &    0.24\\
21 17 &    0.24 &   0.08 &   0.00 &   0.13 &   0.31 &   0.00 &   8.22 &  -3.14 &    6.89 &    0.00\\
22 17 &    0.22 &   0.11 &   0.01 &   0.15 &   0.75 &   8.27 &   8.27 &  -3.12 &    8.53 &    0.00\\
23 17 &    0.21 &   0.09 &   0.04 &   0.10 &   0.60 &   8.36 &   8.31 &  -3.13 &   10.40 &    0.00\\
14 18 &    0.18 &   0.08 &   0.01 &   0.06 &   0.33 &   8.36 &   8.39 &  -3.01 &    6.86 &    0.27\\
15 18 &    0.35 &   0.12 &   0.04 &   0.09 &   0.53 &   8.37 &   8.35 &  -2.98 &    8.18 &    0.18\\
16 18 &    0.54 &   0.12 &   0.05 &   0.14 &   0.86 &   8.43 &   8.30 &  -3.02 &    8.89 &    0.30\\
17 18 &    0.67 &   0.12 &   0.09 &   0.19 &   1.20 &   8.44 &   8.31 &  -3.07 &    9.12 &    0.24\\
18 18 &    0.60 &   0.10 &   0.09 &   0.23 &   4.74 &   8.45 &   8.34 &  -3.35 &    7.95 &    0.39\\
19 18 &    0.46 &   0.08 &   0.04 &   0.19 &   4.26 &   8.41 &   8.32 &  -3.49 &    6.07 &    0.39\\
20 18 &    0.32 &   0.10 &   0.02 &   0.16 &   4.21 &   8.36 &   8.34 &  -3.56 &    4.61 &    0.34\\
21 18 &    0.26 &   0.09 &   0.00 &   0.12 &   0.69 &   0.00 &   8.31 &  -3.30 &    4.96 &    0.06\\
22 18 &    0.23 &   0.10 &   0.01 &   0.16 &   0.93 &   8.39 &   8.39 &  -3.11 &    9.56 &    0.17\\
23 18 &    0.24 &   0.07 &   0.06 &   0.13 &   0.66 &   8.48 &   8.46 &  -3.15 &   12.77 &    0.00\\
24 18 &    0.20 &   0.05 &   0.08 &   0.08 &   0.57 &   8.58 &   8.52 &  -3.10 &   11.79 &    0.00\\
14 19 &    0.21 &   0.10 &   0.02 &   0.07 &   0.28 &   8.39 &   8.41 &  -3.15 &    7.63 &    0.00\\
15 19 &    0.36 &   0.16 &   0.06 &   0.16 &   0.60 &   8.43 &   8.38 &  -3.16 &    7.87 &    0.00\\
16 19 &    0.53 &   0.17 &   0.08 &   0.24 &   1.12 &   8.46 &   8.34 &  -3.17 &    8.11 &    0.15\\
17 19 &    0.68 &   0.17 &   0.12 &   0.26 &   1.60 &   8.46 &   8.31 &  -3.06 &    7.64 &    0.21\\
18 19 &    0.67 &   0.15 &   0.13 &   0.28 &   3.52 &   8.48 &   8.31 &  -3.28 &    6.24 &    0.44\\
19 19 &    0.63 &   0.13 &   0.14 &   0.26 &   3.00 &   8.47 &   8.29 &  -3.36 &    5.00 &    0.45\\
20 19 &    0.49 &   0.12 &   0.09 &   0.24 &   2.91 &   8.45 &   8.35 &  -3.38 &    4.16 &    0.43\\
21 19 &    0.36 &   0.09 &   0.03 &   0.20 &   1.08 &   8.40 &   8.36 &  -3.20 &    4.29 &    0.25\\
22 19 &    0.28 &   0.09 &   0.04 &   0.19 &   1.25 &   8.44 &   8.43 &  -3.15 &    8.86 &    0.33\\
23 19 &    0.27 &   0.07 &   0.08 &   0.16 &   0.92 &   8.52 &   8.46 &  -3.09 &   14.10 &    0.12\\
24 19 &    0.24 &   0.06 &   0.10 &   0.10 &   0.79 &   8.65 &   8.51 &  -3.04 &   14.02 &    0.00\\
12 20 &    0.17 &   0.04 &   0.01 &   0.04 &   0.69 &   8.17 &   8.50 &  -3.05 &    8.36 &    0.00\\
13 20 &    0.16 &   0.06 &   0.01 &   0.03 &   0.68 &   8.39 &   8.48 &  -3.14 &    7.20 &    0.00\\
14 20 &    0.21 &   0.10 &   0.07 &   0.04 &   0.78 &   8.51 &   8.44 &  -3.19 &    7.78 &    0.00\\
15 20 &    0.33 &   0.14 &   0.12 &   0.13 &   0.61 &   8.56 &   8.42 &  -3.17 &    5.96 &    0.00\\
16 20 &    0.46 &   0.16 &   0.11 &   0.21 &   1.08 &   8.55 &   8.38 &  -3.19 &    5.83 &    0.15\\
17 20 &    0.70 &   0.18 &   0.15 &   0.24 &   1.59 &   8.52 &   8.30 &  -3.15 &    6.25 &    0.22\\
18 20 &    0.77 &   0.19 &   0.16 &   0.27 &   2.03 &   8.54 &   8.31 &  -3.21 &    5.17 &    0.38\\
19 20 &    0.83 &   0.18 &   0.25 &   0.28 &   2.18 &   8.56 &   8.32 &  -3.28 &    4.25 &    0.48\\
20 20 &    0.68 &   0.15 &   0.20 &   0.29 &   2.32 &   8.55 &   8.37 &  -3.32 &    3.51 &    0.42\\
21 20 &    0.48 &   0.09 &   0.11 &   0.25 &   2.19 &   8.51 &   8.36 &  -3.34 &    3.07 &    0.23\\
22 20 &    0.32 &   0.06 &   0.04 &   0.24 &   1.72 &   8.51 &   8.42 &  -3.27 &    6.07 &    0.29\\ \hline

\end{tabular}
\end{table*}

\newpage

\setcounter{table}{0}
\begin{table*}
\scriptsize

\caption{continued.}
\begin{tabular}{lllllllllll}\hline\hline
Spaxel & H$\alpha$ & H$\beta$ & [NII] & [OIII] & [OII]& Z (N2) & Z (O3N2) &log U & SSFR & E(B--V) \\
 & & & 6586 & 5007 & 3727,29& 12+log(O/H)&12+log(O/H)& [OII]/[OIII]& M$_\odot$/yr/L/L* & [mag]  \\ \hline\hline
23 20 &    0.27 &   0.05 &   0.07 &   0.23 &   1.23 &   8.57 &   8.44 &  -3.24 &   10.33 &    0.21\\
24 20 &    0.26 &   0.05 &   0.10 &   0.16 &   0.95 &   8.63 &   8.51 &  -3.10 &   12.76 &    0.00\\
25 20 &    0.19 &   0.06 &   0.08 &   0.10 &   0.91 &   8.67 &   8.47 &  -3.08 &   10.85 &    0.00\\
13 21 &    0.16 &   0.05 &   0.00 &   0.03 &   0.61 &   0.00 &   8.55 &  -3.32 &    8.09 &    0.00\\
14 21 &    0.20 &   0.06 &   0.07 &   0.02 &   0.72 &   8.59 &   8.41 &  -3.38 &    9.46 &    0.00\\
15 21 &    0.30 &   0.09 &   0.13 &   0.08 &   0.39 &   8.64 &   8.40 &  -3.17 &    7.96 &    0.00\\
16 21 &    0.42 &   0.13 &   0.15 &   0.17 &   0.95 &   8.61 &   8.40 &  -3.14 &    4.46 &    0.00\\
17 21 &    0.87 &   0.22 &   0.24 &   0.18 &   1.56 &   8.56 &   8.38 &  -3.15 &    5.01 &    0.00\\
18 21 &    1.07 &   0.27 &   0.27 &   0.25 &   2.14 &   8.56 &   8.40 &  -3.30 &    4.12 &    0.14\\
19 21 &    1.07 &   0.29 &   0.36 &   0.29 &   2.36 &   8.59 &   8.43 &  -3.30 &    3.59 &    0.38\\
20 21 &    0.79 &   0.20 &   0.28 &   0.35 &   2.64 &   8.62 &   8.46 &  -3.50 &    2.98 &    0.37\\
21 21 &    0.53 &   0.13 &   0.20 &   0.32 &   2.49 &   8.60 &   8.45 &  -3.52 &    2.80 &    0.23\\
22 21 &    0.42 &   0.09 &   0.11 &   0.30 &   2.17 &   8.57 &   8.45 &  -3.66 &    4.11 &    0.19\\
23 21 &    0.35 &   0.08 &   0.10 &   0.28 &   1.61 &   8.57 &   8.39 &  -3.46 &    8.81 &    0.21\\
24 21 &    0.34 &   0.08 &   0.11 &   0.21 &   1.33 &   8.59 &   8.40 &  -3.38 &   13.43 &    0.00\\
25 21 &    0.26 &   0.05 &   0.07 &   0.16 &   1.16 &   8.57 &   8.37 &  -3.06 &   14.03 &    0.00\\
14 22 &    0.18 &   0.07 &   0.06 &   0.07 &   1.02 &   8.64 &   8.38 &  -2.99 &    8.49 &    0.00\\
15 22 &    0.21 &   0.10 &   0.09 &   0.13 &   0.52 &   8.64 &   8.36 &  -2.78 &    6.39 &    0.00\\
16 22 &    0.43 &   0.19 &   0.15 &   0.29 &   0.72 &   8.61 &   8.36 &  -2.92 &    4.31 &    0.00\\
17 22 &    0.88 &   0.30 &   0.21 &   0.33 &   1.39 &   8.57 &   8.33 &  -3.13 &    4.48 &    0.00\\
18 22 &    1.21 &   0.35 &   0.29 &   0.35 &   2.25 &   8.56 &   8.37 &  -3.37 &    4.23 &    0.00\\
19 22 &    1.21 &   0.38 &   0.35 &   0.32 &   2.60 &   8.59 &   8.42 &  -3.41 &    3.31 &    0.19\\
20 22 &    1.00 &   0.28 &   0.37 &   0.36 &   2.70 &   8.62 &   8.46 &  -3.60 &    3.57 &    0.26\\
21 22 &    0.70 &   0.22 &   0.27 &   0.34 &   2.57 &   8.62 &   8.44 &  -3.58 &    3.47 &    0.33\\
22 22 &    0.51 &   0.14 &   0.16 &   0.28 &   2.29 &   8.60 &   8.41 &  -3.77 &    4.01 &    0.29\\
23 22 &    0.36 &   0.13 &   0.09 &   0.24 &   2.03 &   8.58 &   8.36 &  -3.60 &    5.51 &    0.25\\
24 22 &    0.34 &   0.11 &   0.09 &   0.16 &   1.53 &   8.48 &   8.37 &  -3.45 &   10.56 &    0.00\\
25 22 &    0.26 &   0.08 &   0.05 &   0.14 &   1.26 &   8.43 &   8.37 &  -3.08 &   11.98 &    0.00\\
14 23 &    0.22 &   0.06 &   0.07 &   0.10 &   0.44 &   8.53 &   8.26 &  -2.59 &    8.86 &    0.00\\
15 23 &    0.29 &   0.13 &   0.08 &   0.17 &   0.53 &   8.55 &   8.28 &  -2.88 &    8.08 &    0.01\\
16 23 &    0.50 &   0.23 &   0.14 &   0.35 &   0.64 &   8.54 &   8.30 &  -2.97 &    5.62 &    0.08\\
17 23 &    0.95 &   0.39 &   0.21 &   0.43 &   1.23 &   8.54 &   8.35 &  -3.15 &    5.20 &    0.08\\
18 23 &    1.23 &   0.38 &   0.30 &   0.50 &   2.03 &   8.54 &   8.37 &  -3.35 &    4.58 &    0.00\\
19 23 &    1.20 &   0.42 &   0.29 &   0.47 &   2.56 &   8.56 &   8.41 &  -3.40 &    3.53 &    0.24\\
20 23 &    0.95 &   0.30 &   0.28 &   0.47 &   2.64 &   8.57 &   8.43 &  -3.74 &    3.42 &    0.14\\
21 23 &    0.68 &   0.28 &   0.18 &   0.40 &   2.50 &   8.57 &   8.42 &  -3.72 &    3.36 &    0.15\\
22 23 &    0.53 &   0.17 &   0.16 &   0.28 &   2.33 &   8.54 &   8.34 &  -3.92 &    3.94 &    0.10\\
23 23 &    0.35 &   0.14 &   0.08 &   0.19 &   2.02 &   8.51 &   8.29 &  -3.47 &    4.93 &    0.10\\
24 23 &    0.30 &   0.11 &   0.07 &   0.12 &   1.49 &   8.41 &   8.28 &  -3.34 &    7.70 &    0.00\\
25 23 &    0.20 &   0.08 &   0.00 &   0.11 &   1.03 &   8.31 &   8.31 &  -2.95 &    8.51 &    0.00\\
14 24 &    0.18 &   0.08 &   0.03 &   0.13 &   0.43 &   8.50 &   8.37 &  -2.69 &    6.29 &    0.00\\
15 24 &    0.23 &   0.14 &   0.03 &   0.19 &   0.59 &   8.49 &   8.36 &  -3.03 &    4.98 &    0.08\\
16 24 &    0.35 &   0.22 &   0.04 &   0.38 &   0.57 &   8.50 &   8.31 &  -3.13 &    4.43 &    0.08\\
17 24 &    0.59 &   0.35 &   0.12 &   0.48 &   1.00 &   8.51 &   8.32 &  -3.26 &    4.10 &    0.05\\
18 24 &    0.85 &   0.35 &   0.21 &   0.56 &   1.64 &   8.54 &   8.32 &  -3.34 &    4.05 &    0.00\\
19 24 &    1.05 &   0.41 &   0.22 &   0.58 &   2.24 &   8.53 &   8.37 &  -3.35 &    4.26 &    0.06\\
20 24 &    1.00 &   0.31 &   0.21 &   0.60 &   2.33 &   8.50 &   8.36 &  -3.53 &    4.48 &    0.00\\
21 24 &    0.75 &   0.27 &   0.12 &   0.46 &   2.51 &   8.47 &   8.34 &  -3.49 &    3.93 &    0.00\\
22 24 &    0.45 &   0.13 &   0.09 &   0.30 &   2.32 &   8.44 &   8.28 &  -3.56 &    3.26 &    0.09\\
23 24 &    0.23 &   0.10 &   0.03 &   0.13 &   1.90 &   8.39 &   8.25 &  -3.24 &    3.05 &    0.09\\
24 24 &    0.17 &   0.07 &   0.01 &   0.11 &   1.08 &   8.29 &   8.23 &  -3.17 &    4.80 &    0.00\\
14 25 &    0.19 &   0.04 &   0.03 &   0.10 &   0.06 &   8.52 &   8.38 &  -2.72 &    7.31 &    0.00\\
15 25 &    0.24 &   0.08 &   0.05 &   0.13 &   0.24 &   8.46 &   8.39 &  -3.04 &    5.78 &    0.00\\
16 25 &    0.29 &   0.10 &   0.03 &   0.24 &   0.42 &   8.47 &   8.34 &  -3.17 &    4.49 &    0.00\\
17 25 &    0.48 &   0.23 &   0.14 &   0.31 &   0.73 &   8.49 &   8.39 &  -3.43 &    4.39 &    0.00\\
18 25 &    0.60 &   0.27 &   0.16 &   0.45 &   1.07 &   8.53 &   8.34 &  -3.33 &    4.14 &    0.00\\
19 25 &    0.79 &   0.37 &   0.18 &   0.54 &   1.60 &   8.50 &   8.37 &  -3.25 &    4.76 &    0.05\\
20 25 &    0.78 &   0.32 &   0.11 &   0.64 &   1.94 &   8.46 &   8.29 &  -3.35 &    4.53 &    0.18\\
21 25 &    0.65 &   0.26 &   0.07 &   0.47 &   2.24 &   8.42 &   8.29 &  -3.35 &    4.21 &    0.34\\
22 25 &    0.41 &   0.12 &   0.04 &   0.32 &   2.22 &   8.36 &   8.22 &  -3.32 &    3.40 &    0.39\\
23 25 &    0.23 &   0.04 &   0.02 &   0.11 &   1.56 &   8.30 &   8.21 &  -3.01 &    4.51 &    0.36\\
24 25 &    0.17 &   0.03 &   0.01 &   0.11 &   0.92 &   8.14 &   8.16 &  -2.86 &    6.26 &    0.28\\
25 25 &    0.15 &   0.03 &   0.02 &   0.08 &   0.43 &   8.14 &   8.28 &  -2.68 &    7.93 &    0.00\\
17 26 &    0.26 &   0.12 &   0.05 &   0.23 &   0.50 &   8.47 &   8.37 &  -3.54 &    3.10 &    0.00\\
18 26 &    0.40 &   0.18 &   0.07 &   0.27 &   0.67 &   8.50 &   8.32 &  -3.47 &    4.13 &    0.00\\
19 26 &    0.62 &   0.30 &   0.13 &   0.40 &   1.09 &   8.48 &   8.34 &  -3.25 &    5.65 &    0.06\\
20 26 &    0.74 &   0.30 &   0.10 &   0.54 &   1.38 &   8.46 &   8.28 &  -3.18 &    6.46 &    0.20\\
21 26 &    0.73 &   0.23 &   0.12 &   0.51 &   1.91 &   8.40 &   8.25 &  -3.18 &    7.10 &    0.44\\
22 26 &    0.53 &   0.10 &   0.07 &   0.39 &   1.94 &   8.32 &   8.23 &  -3.18 &    6.88 &    0.52\\
23 26 &    0.30 &   0.02 &   0.02 &   0.18 &   1.46 &   8.24 &   8.39 &  -2.98 &    6.49 &    0.55\\
24 26 &    0.19 &   0.02 &   0.00 &   0.13 &   0.81 &   0.00 &   8.51 &  -2.85 &    7.65 &    0.60\\
17 27 &    0.16 &   0.08 &   0.01 &   0.14 &   0.25 &   8.36 &   8.41 &  -3.44 &    3.08 &    0.00\\
18 27 &    0.25 &   0.13 &   0.04 &   0.19 &   0.29 &   8.36 &   8.36 &  -3.40 &    4.43 &    0.00\\
19 27 &    0.39 &   0.22 &   0.08 &   0.29 &   0.57 &   8.37 &   8.34 &  -3.27 &    5.99 &    0.00\\
20 27 &    0.58 &   0.25 &   0.10 &   0.42 &   0.82 &   8.39 &   8.30 &  -3.22 &    7.59 &    0.00\\
21 27 &    0.67 &   0.20 &   0.12 &   0.45 &   1.32 &   8.41 &   8.24 &  -3.16 &    8.90 &    0.10\\
22 27 &    0.61 &   0.12 &   0.08 &   0.36 &   1.59 &   8.37 &   8.25 &  -3.18 &    9.25 &    0.29\\
23 27 &    0.39 &   0.04 &   0.01 &   0.22 &   1.51 &   8.23 &   8.39 &  -2.97 &    9.67 &    0.58\\
24 27 &    0.23 &   0.03 &   0.00 &   0.13 &   1.18 &   0.00 &   8.64 &  -2.87 &    8.32 &    0.75\\ \hline\hline
\end{tabular}
\label{table:VIMOSfluxtable}
\end{table*}

\newpage

\begin{table*}
\scriptsize
\caption{Properties derived from emission line fluxes from both slits of the HIRES data as plotted in Fig.\ref{HIRESprop}, \ref{060505:revisedrotcurve} and \ref{060505:FWHMplot}. Note that the centre of the H$\alpha$ line is as measured from the data and uncorrected for inclination (this has been later done for the fit of the rotation curve).}

\begin{tabular}{lccccccc}\hline\hline
pos & centre H$\alpha$&FWHM H$\alpha$ & FWHM [OIII] 5007 & Z (N2) & Z (O3N2)  & log [SII]/H$\alpha$ & E(B--V) \\
 arcsec & km s$^{-1}$&km s$^{-1}$&km s$^{-1}$ & 12+log(O/H)  & 12+log(O/H) &    &  [mag] \\ \hline\hline
{\bf Slit1}&&&&&&\\
   0.0&  21.7$\pm$0.4&39.76$\pm$   1.03&  ---&   8.41$\pm$   0.01&   ---&    -0.82&   0.59\\
   0.5&  37.2$\pm$0.6&74.97$\pm$   1.97&  41.88$\pm$   3.49&   8.42$\pm$   0.01&   8.43$\pm$   0.01&   -0.65&   0.02\\
   1.0&  73.3$\pm$0.7&94.32$\pm$   2.28&  68.83$\pm$   5.52&   8.46$\pm$   0.01&   8.46$\pm$   0.01&    -0.58&   0.00\\
   1.5&  112.7$\pm$1.0&71.49$\pm$   2.51& 101.16$\pm$  10.22&   8.47$\pm$   0.01&   8.45$\pm$   0.01&    -0.57&   0.00\\
   2.0&  129.3$\pm$0.4&54.46$\pm$   1.06&  84.54$\pm$   9.55&   8.40$\pm$   0.01&   8.43$\pm$  0.01&    -0.51&   0.00\\
   2.5&  141.5$\pm$0.4&49.53$\pm$   0.93&  52.89$\pm$   5.75&   8.39$\pm$   0.01&   8.40$\pm$  0.01&    -0.55&   0.00\\
   3.0&  146.7$\pm$0.3&49.26$\pm$   0.73&  47.26$\pm$   2.56&   8.36$\pm$   0.01&   8.35$\pm$  0.01&   -0.60&   0.00\\
   3.5&  148.7$\pm$0.3&53.68$\pm$   0.85&  43.28$\pm$   1.56&   8.32$\pm$   0.01&   8.29$\pm$  0.01&    -0.61&   0.00\\
   4.0&  155.7$\pm$0.5&46.30$\pm$   1.30&  46.81$\pm$   1.60&   8.32$\pm$   0.02&   8.29$\pm$  0.01&    -0.68&   0.00\\
   4.5&  161.8$\pm$0.5&40.18$\pm$   1.25&  30.33$\pm$   1.41&   ---&   ---&     ---&   0.00\\
   5.0&  166.9$\pm$0.6&42.46$\pm$   1.55&  35.60$\pm$   1.96&   ---&   ---&   ---&   0.00\\
   5.5&  169.2$\pm$0.9&44.28$\pm$   2.22&  47.15$\pm$   3.33&  ---&  ---&     ---&   0.00\\
   6.0&  175.5$\pm$0.9&43.08$\pm$   2.32&  37.13$\pm$   4.09&   ---&  ---&     ---&   0.00\\
\hline
{\bf Slit2}&&&&&&&\\
   0.0&  111.4$\pm$2.4&48.76$\pm$   6.06& -10.00$\pm$   0.00&   ---&    ---&      ---&   0.00\\
   0.5&  121.3$\pm$1.0&52.06$\pm$   2.45&  51.61$\pm$   8.43&   ---&    ---&     ---&   0.89\\
   1.0&  135.84$\pm$1.0&44.39$\pm$   2.57&  75.22$\pm$   8.19&   ---&    ---&     ---&   0.00\\
   1.5&  147.5$\pm$2.1&62.13$\pm$   5.32&  57.92$\pm$   7.44&  ---&    ---&    ---&   0.06\\
   2.0&  150.6$\pm$2.0&56.62$\pm$   4.90&  37.40$\pm$   4.82&  ---&    ---&     ---&   0.56\\
   2.5&  156.4$\pm$0.7&36.88$\pm$   1.77&  21.99$\pm$   2.91&   ---&    ---&     ---&   0.00\\
   3.0&  164.3$\pm$0.4&43.98$\pm$   1.07&  26.16$\pm$   1.19&   ---&    ---&    ---&   0.85\\
   3.5&  173.3$\pm$0.4&45.95$\pm$   0.94&  33.64$\pm$   1.15&   ---&    ---&      ---&   0.00\\
   4.0&  177.7$\pm$0.6&42.75$\pm$   1.53&  48.14$\pm$   2.21&   ---&    ---&    ---&   0.00\\
   4.5&  181.6$\pm$1.7&51.80$\pm$   4.28&  50.34$\pm$   4.23&   ---&    ---&     ---&   0.00\\
   5.0&  183.4$\pm$3.7&55.17$\pm$   9.06&  19.84$\pm$   2.66&   ---&    ---&    ---&   0.22\\
\hline\hline
\label{table:HIRESval}
\end{tabular}
\end{table*}

\begin{table*}
\caption{Fluxes measured in the integrated spectra of the concentric ellipses extracted from the VIMOS dataset (see Sect.\ref{comparison}) in units of 10$^{-16}$ erg\,cm$^{-2}$\,s$^{-1}$. } 
\begin{tabular}{l|llllll}\hline
distance & [O\sc{ii}] & H$\beta$ & [O\sc{iii}] & [O\sc{iii}] & H$\alpha$ & [N\sc{ii}] \\
 arcsec & $\lambda$3727/29 & $\lambda$ 4862 & $\lambda$ 4960 & $\lambda$ 5008 & $\lambda$ 6564 & $\lambda$ 6585  \\ \hline\hline
0  &   5.42 $\pm$   0.18 &   1.94 $\pm$   0.48  &   1.16 $\pm$   0.10  &   1.34 $\pm$   0.10 &   6.94 $\pm$   1.26 &   0.90 $\pm$   0.22\\
0.8  &   2.37 $\pm$   0.28 &   0.52 $\pm$   0.12  &   0.43 $\pm$   0.07  &   0.90 $\pm$   0.13 &   1.74 $\pm$   0.34 &   0.33 $\pm$   0.14\\
1.6  &   9.28 $\pm$   0.62 &   2.42 $\pm$   0.18  &   0.68 $\pm$   0.08  &   3.98 $\pm$   0.24 &   9.51 $\pm$   1.08 &   2.86 $\pm$   0.32\\
2.4  &  12.60 $\pm$   2.24 &   2.64 $\pm$   0.32  &   1.74 $\pm$   0.36  &   5.70 $\pm$   0.40 &  11.58 $\pm$   1.50 &   2.98 $\pm$   0.44\\
3.2  &  10.85 $\pm$   0.45 &   1.34 $\pm$   0.14  &   1.82 $\pm$   0.18  &   5.02 $\pm$   0.10 &  10.01 $\pm$   1.24 &   2.32 $\pm$   0.24\\
4.0  &  13.23 $\pm$   2.44 &   2.28 $\pm$   0.26  &   1.14 $\pm$   0.24  &   4.12 $\pm$   0.36 &   8.69 $\pm$   1.50 &   1.24 $\pm$   0.28\\
4.8  &  10.92 $\pm$   0.76 &   1.55 $\pm$   0.18  &   1.21 $\pm$   0.10  &   4.24 $\pm$   0.18 &   7.33 $\pm$   0.74 &   1.06 $\pm$   0.10\\
5.6  &   1.86 $\pm$   0.17 &   3.07 $\pm$   0.12  &   1.40 $\pm$   0.12  &   3.84 $\pm$   0.14 &   7.70 $\pm$   0.68 &   1.62 $\pm$   0.18\\\hline\hline
\end{tabular}
\label{ellipses:fluxtable}
\end{table*}

\begin{table*}
\caption{Properties derived from integrated spectra of the concentric ellipses extracted from the VIMOS dataset (see Fig.\ref{ellipses_prop} and Sect.\ref{comparison}). Errors for the metallicity only includes the errors from the emission line fitting which are always smaller than the errors of the different calibrator (0.16\,dex for the N2 and 0.18\,dex for O3N2).} 
\begin{tabular}{llllllll}\hline
 distance & 12+log(O/H) & 12+log(O/H) & log U&  E(B--V) & SFR & SSFR & EW H$\alpha$ \\
 arcsec & (N2) & (O3N2) & & [mag] & M$_\odot$/yr & M$_\odot$/yr/L/L*/spaxel & \AA{}\\ \hline\hline 
0 &   8.33 $\pm$   0.09 &    8.38 $\pm$   0.02 &  -3.50 $\pm$   0.05 &     0.25 &  0.011$\pm$  0.002 &   0.38 $\pm$   0.07 &     -45 $\pm$    5.3\\
0.8 &   8.41 $\pm$   0.12 &    8.33 $\pm$   0.02 &  -3.36 $\pm$   0.12 &     0.18 &  0.003$\pm$  0.001 &   0.21 $\pm$   0.04 &     -22 $\pm$    4.1\\
1.6 &   8.50 $\pm$   0.05 &    8.38 $\pm$   0.02 &  -3.31 $\pm$   0.06 &     0.34 &  0.014$\pm$  0.002 &   0.33 $\pm$   0.04 &     -32 $\pm$    4.1\\
2.4 &   8.47 $\pm$   0.06 &    8.34 $\pm$   0.03 &  -3.30 $\pm$   0.11 &     0.45 &  0.018$\pm$  0.002 &   0.42 $\pm$   0.05 &     -45 $\pm$    5.3\\
3.2 &   8.45 $\pm$   0.05 &    8.27 $\pm$   0.02 &  -3.29 $\pm$   0.03 &     0.96 &  0.015$\pm$  0.002 &   0.49 $\pm$   0.06 &     -41 $\pm$    5.0\\
4.0 &   8.35 $\pm$   0.08 &    8.30 $\pm$   0.03 &  -3.43 $\pm$   0.12 &     0.31 &  0.013$\pm$  0.002 &   0.74 $\pm$   0.13 &     -50 $\pm$    6.0\\
4.8 &   8.35 $\pm$   0.04 &    8.26 $\pm$   0.02 &  -3.35 $\pm$   0.05 &     0.51 &  0.011$\pm$  0.001 &   0.68 $\pm$   0.07 &     -47 $\pm$    5.9\\
5.6 &   8.43 $\pm$   0.04 &    8.37 $\pm$   0.01 &  -2.77 $\pm$   0.06 &     0.00 &  0.012$\pm$  0.001 &   0.88 $\pm$   0.08 &     -65 $\pm$    7.4\\

\hline\hline
\end{tabular}

 \label{ellipses:comptable}
\end{table*}
\vspace{10cm}

\label{lastpage}

\end{document}